\documentclass[article,twocolumn,nofootinbib,showkeys]{revtex4-1} 
\usepackage[utf8]{inputenc}
\usepackage[T1]{fontenc} 
\usepackage[brazil,brazilian,english]{babel}  
\usepackage{amsfonts,amstext,amssymb,amsmath}
\usepackage{bbold} 
\usepackage{graphicx} 
\usepackage{color,xcolor} 
\usepackage{pygtex}
\usepackage{relsize}
\usepackage{algorithm,algpseudocode,algorithmicx} 
\usepackage{hyperref}
\hypersetup{
    colorlinks,%
    citecolor=blue,%
    filecolor=blue,%
    linkcolor=blue,%
    urlcolor=blue
}

\newenvironment{abstractpage}
{}

\begin{document}

\title{Pedagogical introduction to equilibrium Green's functions: \\
condensed-matter examples with numerical implementations \\ \vspace{0.3cm} 
\small Introdução pedagógica às funções de Green de equilíbrio: exemplos em matéria 
condensada \\ com implementações numéricas}

\author{Mariana M. Odashima, Beatriz G. Prado and E. Vernek}
\affiliation{Instituto de Física, Universidade Federal  de Uberlândia, 
Uberlândia-MG, 38400-902, Brasil}

\begin{abstract}
The Green's function  method has applications in several fields in Physics, from 
classical differential equations to quantum many-body problems. In the quantum 
context, Green's functions are correlation functions, from which it is possible 
to extract information from the system under study, such as the density of 
states, relaxation times and response functions. Despite its power and 
versatility, it is known as a laborious and sometimes cumbersome method. Here we 
introduce the equilibrium Green's functions and the equation-of-motion 
technique, exemplifying the method in discrete lattices of non-interacting 
electrons. We start with simple models, such as the two-site molecule, the 
infinite and semi-infinite one-dimensional chains, and the two-dimensional 
ladder. Numerical implementations are developed via the recursive Green's 
function, implemented in \texttt{Julia}, an open-source, efficient and 
easy-to-learn scientific language. We also present a new variation of the 
surface recursive Green's function method, which can be of interest when 
simulating simultaneously the properties of surface and bulk.
\begin{abstractpage}
\mbox{}
\mbox{}\\ 
\vspace{0.5cm}

 O método das funções de Green possui  aplicações em diversos campos da Física, 
desde equações diferenciais clássicas a problemas quânticos de muitos corpos. No 
contexto quântico, as funções de Green são funções de correlação, das quais é 
possível extrair informação sobre o sistema em estudo, tais como densidade de 
estados, tempos de relaxação e funções respostas. Apesar de seu poder e 
versatilidade, este método é conhecido por ser trabalhoso e às vezes intrincado. 
Neste trabalho introduzimos as funções de Green de equilíbrio e a técnica de 
equação de movimento, exemplificando o método em redes discretas de elétrons 
não-interagentes. Começamos com modelos simples, como a molécula de dois sítios, 
as cadeias unidimensionais infinita e semi-infinita, e a rede escada em duas 
dimensões. Implementações numéricas são desenvolvidas através das funções de 
Green recursivas, implementadas em \texttt{Julia}, uma linguagem científica de 
código aberto, eficiente e de fácil aprendizado. Também apresentamos uma nova 
variante do método de função de Green recursiva de superfície, que pode ser útil 
para simular simultaneamente as propriedades de superfície e bulk. \\
\footnotesize\textbf{Palavras-chave: }{Funções  de Green; Transporte quântico; 
Física de baixa dimensionalidade; Tight-binding; Densidade de estados}
\end{abstractpage}
\end{abstract}

\maketitle

\section{Introduction}

The Green's functions method is a powerful mathematical tool to solve linear 
differential equations. These functions were named after the English miller, 
physicist and mathematician George Green (1793-1841) 
\cite{CannellBio2001,revCannell2004,Cannell1993}. His seminal work ``An essay on 
the application of mathematical analysis to the theories of electricity and 
magnetism'' (1828) \cite{GreenEssay} developed a theory of partial differential 
equations with general boundary conditions, introducing the so-called Green's 
theorem (also known today as Green's second identity\footnote{Today, Green's 
identities are a set of three vector equations relating the bulk with the 
boundary of a region on which differential operators act, closely related to 
Gauss' divergence and Stokes' curl theorems. Green's second identity allows the 
conversion of a triple integral of laplacians within a volume into a double 
integral of gradients over its surface boundary: \begin{equation} \int_\Omega 
\left( u \nabla^2 v - v\nabla^2 u  \right) dv = \int_S \left( u \frac{\partial 
v}{\partial n} - v \frac{\partial n}{\partial n} \right) ds \,, \end{equation} 
where $\Omega$ is a volume bounded by the closed surface $S$, where $ 
\frac{\partial u}{\partial n}= \vec n \cdot \nabla u$, $\vec n$ is the outward 
normal to the boundary $S$ with unit length. This formula holds for regular 
functions $u$ and $v$ defined in $\Omega$. }), and the Green's functions 
\cite{Grattan, Jahnke, TriggOhtaka}. This essay was self-published by Green for 
private distribution among family and friends, and was later rediscovered by 
Lord Kelvin, being examined by Sturm, Liouville, Dirichlet, Riemann, Neumann, 
Maxwell, and others \cite{Lindstrom}. The Green's functions were born as 
\textit{auxiliary} functions for solving boundary-value problems. The latter are 
differential equations with constraining boundary conditions, which specify 
values that the solution or its normal derivative take on the boundary of the 
domain. Boundary-value problems arise in several problems in 
physics, for instance, heat conduction in solid bodies, described by the 
diffusion or heat conduction equation: $\frac{\partial^2 \varphi}{\partial x^2} 
- \frac{1}{k}\frac{\partial \varphi}{\partial t} =0$; in charge distributions in
surfaces by using the Poisson equation: $\nabla^2 \varphi = -\rho/\epsilon_0$;
vibration in strings and membranes and wave propagation along 
special geometries, described by: $\frac{\partial^2 \varphi}{\partial x^2} - 
\frac{1}{v^2} \frac{\partial^2 \varphi}{\partial t^2} =0$, the wave equation 
\cite{Butkov, Carmen, Griffiths}.
From the Green's functions, a whole theory of partial differential equations 
arised, paving the way for the development of functional analysis, the 
branch of mathematics dedicated to the infinite-dimensional vector 
spaces and operators. By the end of the XIX century many boundary-value problems 
were approached in acoustics, hydrodynamics, thermodynamics and electromagnetism. 
Before we examine the development of the Green's functions in quantum mechanics, we 
shall review some of the general properties of a Green's function.

\subsection{Classical Green's functions}

Formally, a Green's function is a solution of a \textit{linear differential equation} 
with a \textit{Dirac delta} \textit{inhomogeneous source} (sometimes referred as a 
delta or unit pulse) with \textit{homogeneous boundary conditions}. 
Let us clarify the emphazised concepts. A differential equation is said to be 
\textit{linear} if the function $f(x)$ and all its derivatives $f^{(n)}(x)$, 
$n=(1,2, \cdots, n)$, appear linearly. There is no product of the function and 
its derivatives, such as $f(x).f^{''}(x)$, and no powers of the function or of its 
derivatives beyond the first power. For example, in an ordinary differential 
equation, it should read:
\begin{equation} a_1(x)f(x) + a_2(x)f'(x) + \cdots + 
a_n(x)f^{(n)}(x) = g(x) \,, \label{eq:linearedo}
\end{equation} 
On the other hand, the coefficients 
$a_n(x)$ are arbitrary differentiable functions. Linearity of the operators is 
essential for the validity of the superposition principle, that allows 
the linear combination of solutions.

If a differential equation has a term on the right-hand-side 
(r.h.s.) of the equation that does not depend on your function $f(x)$,
we classify it as an \textit{inhomogeneous} differential equation.
For example, in Eq.~\eqref{eq:linearedo}, a linear \textit{homogeneous}
differential equation would have $g(x)=0$, and an \textit{inhomogeneous one} 
would have a non-zero function $g(x)$ on the r.h.s., or a non-zero constant $c$.

The differential equation we will be concerned with has a special inhomogeneity 
function, the \textit{Dirac delta} $\delta(x-x')$. Put simply, this object is 
defined to be zero when $x\neq x'$, and infinite at $x=x'$:
\begin{align}
 \delta(x-x')= \left\{ 
 \begin{aligned} &\infty, \quad \textrm{at } x=x' \\
		 &0, \quad \textrm{otherwise} \,.
  \end{aligned}
 \right.
\end{align}
Rigorously, the Dirac delta is not a function, since it would require 
to have a definite value for each point in 
its domain, but is instead classified as a distribution. Its most important property is
\begin{equation}
f(x) = \int_{-\infty}^{\infty} \delta(x-x')f(x') dx' \,, \label{eq:Dirac}\end{equation} 
where $f(x)$ is any continuous function of $x$. For other interesting 
properties of the Dirac delta, please check Refs. 
\cite{Butkov,Carmen, DiracP}.

Lastly, one often needs to impose boundary conditions on the solutions, 
meaning, conditions on the function or on its derivative at the boundary of the domain.
If their values are zero, we call them \textit{homogeneous boundary} conditions. 
For example, for a 
function $f(x)$ with boundary at $x=L$, homogeneous boundary conditions would 
correspond to $f(x=L)=0$ or $f'(x=L)=0$.

Now shall we return to the classical Green's functions. 
To put the mathematical problem in perspective, imagine one would like to  solve 
a partial linear inhomogeneous differential equation, say, 
\begin{equation}
 D f(x)=g(x) \,,
\end{equation}
where $D$ a linear differential operator, $f(x)$ is the desired solution, and 
$g(x)$ the inhomogeneity source. 

The \textit{particular} solution $f(x)$ can be formally found with the aid of a 
function $G(x,x')$:
\begin{equation}
f(x)=\int G(x,x')g(x') dx' \,, \label{eq:GreenInt} 
\end{equation}
where the Green's function $G(x,x')$ is defined as the solution of a differential  
equation with a delta inhomogeneity: 
\begin{equation}
D G(x,x')=\delta(x-x')  \,. \label{eq:deltaG}
\end{equation}

To verify this, act with $D$ on both sides of Eq.~\eqref{eq:GreenInt} 
and make use of the Dirac delta fundamental property, Eq.~\eqref{eq:Dirac}. 
Note that $D$ acts on the $x$ coordinate, keeping $x'$ fixed. 

One can interpret Eq.~\eqref{eq:GreenInt} by considering 
the Green's functions as a ``building block'' for finding the particular 
solution $f(x)$, since they are solutions to delta-impulse equations. 
In signal processing fields, the Green's function is often referred 
to as a response function, connecting a perturbation or ``input signal'' $g(x)$ 
to the ``output'' $f(x)$.

Before turning to applications, we should remark that if one wishes to find 
the complete general solution to $f(x)$, the solution of the homogeneous 
equation $D h(x)=0$ must be added to Eq.~\eqref{eq:GreenInt}, which is 
the particular solution. 
The solution of the homogeneous equation is found by satisfaction of 
inhomogeneous boundary conditions \cite{Carmen}. 

It might be interesting to have an example of how 
Eq.~\eqref{eq:GreenInt}  can work in practice. The equation relating the 
electric potential $\varphi(\mathbf{r})$  to a given charge density distribution 
$\rho(\mathbf{r})$ is Poisson's equation:
\begin{equation}
\nabla^2 \varphi(\mathbf{r}) = -\frac{\rho(\mathbf{r})}{\epsilon_0} \,. 
\end{equation}
The most common boundary condition is requiring that $\varphi(\mathbf{r})$ goes  to 
zero at infinity. 

From Eq.~\eqref{eq:GreenInt}, the potential $\varphi(\mathbf{r})$ can be obtained  
with the help of a Green's function
\begin{equation}
 \varphi(\mathbf{r})=-\frac{1}{\epsilon_0}\int G(\mathbf{r},\mathbf{r}')\rho(\mathbf{r}') d\vec 
r'  \,, \label{eq:GPoisson}
\end{equation}
where $G(\mathbf{r},\mathbf{r}')$ satisfies the inhomogeneous equation
\begin{equation}
 \nabla^2 G(\mathbf{r},\mathbf{r}')=\delta (\mathbf{r} -  \mathbf{r}') \,. \label{eq:Gdelta}
\end{equation}

The solution to Eq.~\eqref{eq:Gdelta} can be identified physically. By 
associating the Dirac delta $\delta(\mathbf{r} - \mathbf{r}')$ with a 
point charge at $\mathbf{r}'$, we can find a corresponding potential. 
Considering a point charge $q=1$ at $\mathbf{r}'$, the electric 
potential is simply
\begin{equation}
  \varphi(\mathbf{r})=\frac{1}{4\pi \epsilon_0} \frac{1}{|\mathbf{r} - \mathbf{r}'|}.  
\label{eq:punctual}
\end{equation}
Removing from \eqref{eq:punctual} the prefactor $-1/\epsilon_0$ of  
Eq.~\eqref{eq:GPoisson}, the appropriate Green's function to the localized 
charge problem of Eq.~\eqref{eq:Gdelta} is
\begin{equation}
  G(\mathbf{r},\mathbf{r}')=-\frac{1}{4\pi} \frac{1}{|\mathbf{r} - \mathbf{r}'|}, \label{eq:Gcharge}
\end{equation}
which satisfies homogeneous boundary conditions, since in the limit of  $|\mathbf{r} 
- \mathbf{r}'| \to \infty$, $G$ goes to zero.

Substituting \eqref{eq:Gcharge} into Eq.~\eqref{eq:GPoisson}, we have  that for 
an arbitrary charge density distribution, the solution of Poisson's equation is 
given by the following integral over space:
\begin{equation}
 \varphi(\mathbf{r})=\frac{1}{4 \pi \epsilon_0}\int \frac{\rho(\mathbf{r}')}{|\mathbf{r}  - 
\mathbf{r}'|} d\mathbf{r}' \,,
\end{equation}
which verifies to be a correct result in electrostatics \cite{Griffiths}.

Although this is a quite simple example, the Green's function technique as 
presented can be applied to other physical problems 
described by \textit{linear} differential equations. During the late 
half of the XIX century, it became a central tool for solving
boundary-value problems. Further examples in vibrations and 
diffusion phenomena, as well as other ways of constructing Green's functions can 
be found in the references \cite{Butkov, Carmen, TriggOhtaka, cole2010heat}. 

\subsection{Quantum Green's functions}

By the beginning of the 20th century, Green's functions were generalized  to the 
theory of linear operators, in particular, they were applied to the class of 
Sturm-Liouville operators  \cite{Sturm}. These are second-order linear 
differential equations that depend linearly on a parameter $\lambda$ as an 
eigenvalue problem: $\mathcal{L}\varphi = \lambda \varphi$. The study of the 
existence of eigenvalues $\lambda$, and of the complete set of eigenfunctions 
$\varphi$ became known as Sturm–Liouville theory. From this set, 
the Green's functions could now be built as a Fourier-like, or \textit{spectral} 
expansion. As a generalized technique, the Green's functions allowed conversion 
of a differential problem into integral operator problems \cite{Lindstrom}.

With the emergence of quantum mechanics, functional analysis and the theory of 
linear operators gained new significance. They are present at the very 
foundations of quantum mechanics, from Hilbert's vector space to Heisenberg's 
matrix formulation, and in Schrödinger's continuous wave mechanics (the 
one-dimensional Schrödinger's equation is one example of a Sturm-Liouville 
problem).

Schrödinger's equation is a celebrated piece of the quantum puzzle that 
tormented early twentieth century physicists. One should remark that the 
Schrödinger equation cannot be rigorously derived from any physical principle; 
it was postulated from Hamilton-Jacobi analogues \cite{Schrodinger} 
describing the propagation of a scalar field, the wave function,  using a 
diffusion equation. As a very historical note, Schrödinger even referenced the 
Green's function in a footnote of one of his 1926 papers \cite{Schrodinger}, 
citing Cornelius Lanczos' work. Lanczos had tried to develop an integral 
representation of Born and Jordan's matrix equations \cite{Taketani}, 
finding the Green's function along his formulation.

Shortly after Schrödinger's first papers, Max Born proposed a wave-mechanical 
model of atomic collisions \cite{Born1926,WheelerZurek}, developing
the probabilistic interpretation of this wave function.
His study was based on the free-particle wave function, a plane wave. In order to 
find the new scattered wavefunction, Born built a perturbation expansion 
in the first power of the potential, starting from the free solution.
This first order is known today as the first Born approximation, generalized 
in the Lippmann-Schwinger equation for scattering \cite{LippmannSchwinger},
presented in Quantum Mechanics courses \cite{SakuraiModern}. In its time-dependent form, 
Schrödinger's equation reads
\begin{equation}
 i\hbar \frac{\partial}{\partial t} \Psi({\bf r}, t) = H \Psi({\bf r}, t) \,. \label{eq:TDSE}
\end{equation}

Eq.\eqref{eq:TDSE} is the quantum nonrelativistic equivalent of second 
Newton's law $\frac{d\mathbf{p}}{dt}=\mathbf{F}$, governing instead the time 
evolution of the wavefunction $\Psi({\bf r}, t)$. 
The right hand side of Eq.\eqref{eq:TDSE} has the Hamiltonian operator,
$H = -\frac{\hbar^2}{2m}\nabla^2 +V({\bf r},t)$, whose expectation value 
is the total energy. The first term is the kinetic energy, rewritten 
using the momentum operator $\mathbf{p}=-i\hbar\nabla$, and the  
second, the external potential. In this formulation, the eigenstates of 
the Hamiltonian play an important role, since their time evolution 
is simple to calculate (i.e. they are stationary). 

The time-dependent Schrödinger equation is a linear partial differential 
equation. Also, it is of first order in time, so an initial condition 
must be specified. Although it is a homogeneous equation, we can 
rearrange the terms as
\begin{equation}
\left[i\hbar \frac{\partial}{\partial t}+\frac{\hbar^2}{2m}\nabla^2\right] 
\Psi({\bf r}, t) =  V({\bf r},t) \Psi({\bf r}, t)\,, \label{eq:SEEDP}
\end{equation}
in order to treat the potential as a source of inhomogeneity. But note that
it is not, since the right-hand-side also depends on the function $\Psi({\bf r}, t)$. 
Ultimately, we will need a recursive solution to find $\Psi({\bf r}, t)$,
or an iterative procedure.
This kind of self-consistent solution is achieved by the 
Lippmann-Schwinger equation for the wave function or a Dyson's equation 
\cite{LippmannSchwinger, SakuraiModern, Bruus, Dyson} for the Green's function.
The basic idea would be to use the free solutions, those in the absence of an 
external potential, to solve the more general problem, with an external potential.

Therefore Eq.\eqref{eq:SEEDP} is where Green's functions come into play. 
Instead of solving Schrödinger's equation for wave functions, one can 
equivalently look for 
the Green's function that solves the inhomogeneous problem
\begin{equation}
 \left[i\hbar \frac{\partial}{\partial t}+\frac{\hbar^2}{2m}\nabla^2\right] 
G({\bf r}, t; {\bf r'}, t')= \delta({\bf r}-{\bf r'})\delta(t-t')  \label{eq:SEGF} \,.
\end{equation}
From the theory of Green's functions we already know that an inhomogeneous 
solution similar to Eq.~\eqref{eq:GreenInt}, may be written as
\begin{equation}
 \Psi({\bf r}, t) = \int G({\bf r}, t; {\bf r'}, t') \Psi({\bf r'}, t') \,d^3r' \,. \label{eq:PsiGF}
\end{equation}
Note the difference with respect to Eq.~\eqref{eq:GreenInt}, where the 
solution itself enters the integral.
The equation above describes the time evolution of the wave function from
a given time and position $({\bf r'},t')$, evolving it to another time and 
space $({\bf r},t)$. This is why the Green's function is known as the 
\textit{propagator}.

In order to give a broader picture the propagating character of the Green's 
function, let us rewrite the wave function in terms of the time evolution 
operator\footnote{For a time-independent 
Hamiltonian and in the Schrödinger picture, the solution to Eq.~\eqref{eq:TDSE} is 
$\Psi({\bf r}, t)= e^{-\frac{i}{\hbar} H(t-t_0)}\Psi({\bf r}, t_0)$. 
Here we see the time-evolution operator $U(t,t')=e^{-\frac{i}{\hbar} H(t-t')}$ 
that evolves the wave function $\Psi({\bf r}, t')$ to $\Psi({\bf r}, t)$ 
in infinitesimal time intervals. It has important properties such as unitarity, 
$U=U^\dagger$, which preserves the norm of the wavefunction.}
$U(t,t')=e^{-\frac{i}{\hbar} H(t-t')}$.
For simplicity, we can represent the wavefunctions as state vectors in the 
position representation\footnote{In the position representation (and Dirac notation), 
the bra $\langle \mathbf{r} |$ is associated to a spatial function base.} 
as $ \Psi({\bf r},t) = \langle \mathbf{r} | \Psi(t) \rangle $.
Writing $\Psi(t)$ as the evolution from $\Psi(t')$, and using the 
closure relation $\int |\mathbf{r'} \rangle \langle \mathbf{r'} | 
d^3r' = \mathbb{1}$,
\begin{align}
 \Psi({\bf r},t) =& \langle \mathbf{r} | \Psi(t) \rangle = \langle \mathbf{r} | 
e^{-\frac{i}{\hbar} H(t-t')} \Psi(t') \rangle \,, \nonumber \\
\Psi({\bf r},t)  =& \int \langle \mathbf{r} | e^{-\frac{i}{\hbar}  H(t-t')} | 
\mathbf{r'} \rangle \langle \mathbf{r'} | \Psi(t') \rangle d^3r' \,. 
\end{align}
which reproduces Eq.~\eqref{eq:PsiGF} if we define (this is not yet our
final definition, we will develop them only for pedagogical purposes),
\begin{equation}
  G({\bf r}, t; {\bf r'}, t') = \langle \mathbf{r} | e^{-\frac{i}{\hbar} 
H(t-t')} | \mathbf{r'} \rangle =  \langle \mathbf{r}, t| \mathbf{r'}, t' \rangle 
\,,
\end{equation}
where $\langle \mathbf{r}, t | =\langle \mathbf{r} | e^{-\frac{i}{\hbar} Ht}  $ 
and $|\mathbf{r'}, t' \rangle= e^{\frac{i}{\hbar}Ht'} | \mathbf{r'} \rangle$. 
Thus, we have associated the Green's function to the probability amplitude of 
finding the particle in a state $\langle \mathbf{r}, t|$ given that it started 
at $|\mathbf{r'}, t' \rangle $. It is interesting to note that Paul Dirac 
\cite{Dirac}, while attempting develop a Lagrangian or path-integral formulation 
of quantum mechanics in the 1930's, found the propagator as the overlap of two 
functions in different positions and times.  

The eigenstates of the Hamiltonian form a complete set, which we cast 
$|n \rangle$ in the vector notation. Inserting again a 
completeness relation, $\sum_n | n \rangle \langle n | = \mathbb{1}$,
\begin{align}
  G({\bf r}, t; {\bf r'}, t') =  \sum_n \langle \mathbf{r} | e^{-\frac{i}{\hbar} 
H(t-t')} | n \rangle \langle n | \mathbf{r'} \rangle \,.
\end{align}

Since $H$ acts on the eigenstate $|n \rangle $, and the projection 
$\langle \mathbf{r} | n  \rangle$ is the eigenfunction $\varphi_n(r)$,
\begin{align}
  G({\bf r}, t; {\bf r'}, t') &=  \sum_n \langle \mathbf{r} | n \rangle \langle 
n | \mathbf{r'} \rangle e^{-\frac{i}{\hbar} E_n(t-t')}  \\
  &= \sum_n \varphi_n(\mathbf{r}) \varphi_n^*(\mathbf{r'})  e^{-\frac{i}{\hbar} 
E_n(t-t')} \, . \label{eq:GEspectral}
\end{align}
This Green's function satisfies Eq.~\eqref{eq:SEGF}.
By Fourier transforming Eq.~\eqref{eq:GEspectral} to energy or frequency 
domain, one obtains a spectral form of the Green's function (again, not 
yet our in final convention):
\begin{align}
  G({\bf r},  {\bf r'}; E) = \sum_n i \frac{\varphi_n({\bf 
r})\varphi^*_n({\bf r'})}{E-E_n}\label{eq:Gspectral} \,,
\end{align}
which has poles at the eigenenergies. Please note that so far, we have 
inspected the quantum Green's functions as propagators, but we have not 
constrained the particle to propagate in a certain direction of time, 
which will be perfomed shortly\footnote{To ensure that particles propagate 
from times $t' < t$, we must correct all equations above with a Heaviside 
function $\theta(t-t')$, and change the analyticity domain, by adding an 
infinitesimal shift $i\eta$ in the denominator of Eq.~\eqref{eq:Gspectral}. 
Later we will return to this point.}.

In the 1950's and 60's the quantum Green's functions were introduced as 
propagators in the quantum field theory by Feynman and Schwinger. 
Feynman \cite{Feynman48,Feynman49} transformed Dirac's 
observations on the quantum propagators into a more rigorous formalism.
He developed the path-integral formalism, interpreting Eq.~\eqref{eq:PsiGF} 
as the sum of the probabilities of the particle taking different individual 
paths. In addition, Feynman invented a graphical form of representing terms
of a perturbation expansion of a scattering formalism, the Feynman diagrams.

At this point we need to switch from the so-called first quantization, from 
Schrödinger's wave mechanics, to quantum fields,  using the technique of 
``second quantization''.  Stating very briefly, the Schrödinger equation 
describes the undulatory behavior of matter, such as electrons, by 
means of wave functions \cite{Lancaster}. But other wave phenomena were shown 
to behave as particles \textit{e.g.}, phonons (lattice vibrations with a 
wavelength) or photons (excitations of the electromagnetic field). The second 
quantization language treats particles and waves as a quantum field. It has 
several advantages over ``first 
quantization'', being more adequate for many-particle physics.

To clarify the definition of a field propagator, let us consider a  thought 
experiment. Imagine that a particle is created in the ground-state of an 
interacting system\footnote{Theoreticians often make this distinction between 
interacting and noninteracting systems. This means that the potential 
$V$ in the Hamiltonian will be present (\textit{e.g.} due to particle 
scattering), or not,
so that we return to the simple free-particle system ($V=0$). 
In practice, the solvable system will be a building block for the more complex 
ones.}.
That particle probes the system, which has its own complex interactions, 
even probably causing excitations, but at the end it is annihilated 
and the system returns to the ground state. In quantum field theory one does not 
deal with wave functions, but instead with one special state, the vacuum  
$|0\rangle$, andthe creation ($c^\dagger_i$) and the annihilation  
($c_i$) operators, where we already assume fermionic fields.
The so-called occupation number representation specifies the number of identical 
particles in each quantum state. 

Feynman introduced a new quantum field propagator. He accounted for the 
propagation of virtual particles and antiparticles, which propagate 
forward and backward in space-time, inserting a 
Wick's time-ordering symbol $T$, that guarantees causal time orderings (we will 
detail the properties of this operator in the following section). 
The Feynman propagator definition reads then 
\cite{Lancaster}:  
\begin{align}
&G(\mathbf{r},t;\mathbf{r'},t')=\big\langle T \big[ \psi(\mathbf{r},t)  
\psi^\dagger(\mathbf{r'},t') \big]   \big\rangle = \label{eq:GFeynman}\\
 &=\theta(t-t') \langle \psi(\mathbf{r},t) \psi^\dagger(\mathbf{r'},t)  
\rangle -\theta(t'-t) \langle  \psi^\dagger(\mathbf{r'},t') \psi(\mathbf{r},t)   
\rangle \,, \nonumber
\end{align}
where the expectation values are evaluated over the interacting ground state 
of the system (later we will generalize to a quantum ensemble of states).
The propagator consists of two parts. In the first, a particle 
is created by $\psi^\dagger(\mathbf{r'},t')$ at position $\mathbf{r'}$ and time 
$t'<t$ and later it is destroyed at the position $\mathbf{r}$ and time $t$. In 
the second part, an antiparticle is created at $\mathbf{r}$ and time $t<t'$ and 
propagates to the position $\mathbf{r'}$, where it is annihilated at time $t'$. 
At this point, we have almost arrived at the many-particle Green's function 
definition that we will adopt. The differences are that expectation values can 
be evaluated in the ground-state or in an ensemble, and we  insert a $-i$ factor 
in our Green's function, in order to avoid the imaginary  factor that appeared 
in Eq.~\eqref{eq:Gspectral}.

Julian Schwinger realized the power of Green's functions in quantum 
field theory. In his very interesting  lecture ``The Greening of Quantum Field 
Theory - George and I'' \cite{Schwinger1993}, Schwinger reviewed Green's 
idea and how it came to post-war developments of quantum field theory, 
finally reaching condensed-matter physicists. One can find several seminal works 
on Green's functions in the condensed-matter literature. To give some examples, 
Martin and Schwinger applied quantum field theory in many-particle physics
\cite{Martin}, introducing the ``lesser'' Green's functions to evaluate particle 
currents and spectral amplitudes, and exploiting the equation-of-motion 
technique with approximate two-particle Green's functions. Kadanoff and Baym 
developed the thermodynamic many-particle Green's function using a 
grand-canonical ensemble average, with periodic boundary conditions along an 
imaginary time axis \cite{Kadanoff, KadanoffBook}, presenting 
conserving approximations and their diagrammatics. Within perturbation theory, 
the Green's functions can be expanded in series and acquires a recursive form, 
known as Dyson's equations \cite{Dyson, Bruus}. 

Due to its versatility, the Green's function method is quite popular in 
many-particle physics. It has also been generalized to particle scattering, 
far from equilibrium physics, finite temperatures, 
statistical mechanics, and other fields. These propagators are naturally 
correlation functions, connecting different positions and times, \textit{e.g.} 
$G({\bf r}_1, t_1;{\bf r}_2,t_2)$.

Nevertheless, due to the arid formalism presented in most of the textbooks, 
the method still scares young students. In view of this, here we aim to 
provide a pedagogical introduction to the Green's functions with practical 
examples. We will be focused on an introductory level of \textit{noninteracting} 
condensed-matter models \textit{i.e.}, without electron-electron Coulomb 
interaction. We will apply the Green's functions to quantum equilibrium 
properties of atomic lattices, described by Hamiltonians in a localized basis 
``tight-binding'' or in an occupation Fock basis, as usually formulated in 
many-particle physics. The fundamentals and definitions can be found for instance, in 
Refs.\cite{Bruus,Mahan,Altland,Ivan}. For fermions, the operator 
ordering is of utmost importance and their algebra should be revised. 
Here we will only add some remarks throughout the text.

\subsection{Electron Green's function}

We will start with formal definitions of the electron Green's function, our object 
of study.
The single particle electron Green's function is defined as the statistical 
expectation value of the product of fermion operators at different 
positions $i$ and $j$ and different times $t$ and 
$t^\prime$. For instance, the so-called ``causal'' Green's function reads
\begin{equation}
G^c_{ij}(t,t')=-i \big\langle T \big[ c_i(t)  c_j^\dagger(t') \big] 
\big\rangle \,, \label{Gcausal}
\end{equation}
where  $c_j^\dagger$ creates and electron at the $j$-th site at time $t'$ 
and  $c_i$ annihilates an electron in the $i$-th at time $t$. We have already 
introduced this causal Green's function in Eq.~\eqref{eq:GFeynman}.
The difference is the imaginary factor $-i$, which Mattuck describes as 
``decorative'' \cite{Mattuck}, and the fermionic creation and annihilation 
operators, expressed in a discrete basis. In this paper we consider atomic units 
in which we set $\hbar=e=m=1$, such that the usual prefactor  $-i/\hbar$ is
simplified. In Eq.~\eqref{Gcausal} we have the time-ordering operator,
\begin{equation}
 T\big[ c_i(t) c_j^\dagger(t')\big] = \theta(t-t^\prime)c_i(t)  c_j^\dagger(t') 
- \theta(t'-t)c_j^\dagger(t')c_i(t)  \,, \label{eq:Torder}
\end{equation}
which guarantees causal orderings. This is due to the properties of 
the Heaviside function $\theta(x)$\footnote{The Heaviside step function $\theta(x)$ 
is defined by \begin{equation}
\theta(x)=\left\{
\begin{array}{r l}
1, & \text{ for } x > 0,\\
0, & \text{ for } x < 0.
\end{array}
\right.
\end{equation} It has a jump discontinuity at $x=0$, for which the value usually taken 
is $1/2$. The derivative of $\theta(x)$ is the Dirac delta $\delta(x)$.}. 
Please verify that in each term of Eq.~\eqref{eq:Torder}, the 
fermionic operator that appears on the left always acts at time later than the right 
one. This rule is known as ``\textit{later to the left}''. Since we are dealing 
with electron operators, we should recall that the operators satisfy the 
anti-commutations relations $\{c_i,c_j\} = 0$, $\{c_i^\dagger,c_j^\dagger\} = 0$ 
and $\{c_i,c_j^\dagger\} = \delta_{ij}$, where the anti-commutator is defined 
as $\{A,B\}=AB+BA$, and the Kronecker function $\delta_{ij}$ assumes the 
values $0$  if $i\neq j$, and $1$ if $i=j$.

Besides the causal Green's function defined above, we introduce two other 
Green's functions from  which many important physical quantities are more easily 
extracted. For example, for times $t > t'$ and $t < t'$, the retarded and advanced 
Green's functions are defined as
\begin{align}
 G_{ij}^r(t,t')&= -i \theta(t-t') \Big\langle \left\{ c_i(t),  
c_j^\dagger(t') \right\} \Big\rangle \,, \label{Gretard}\\
 G_{ij}^a(t,t')&= \;\;-i\theta(t-t') \Big\langle \left\{ c_i(t),  
c_j^\dagger(t') \right\} \Big\rangle \,. \label{Gadvance}
\end{align}
where  $G^r$ is non-zero only for $t > t'$, such that we can calculate the 
response of the system after it has been perturbed. This is why it is 
called retarded Green's function. The advanced Green's function is defined as 
the adjoint of the retarded Green's function, $[G^r]^\dagger=G^a$. This means that, 
having determined one of them, we can immediately calculate the other.

It is important to note that the Green's functions
carry information about the system \textit{excitations}, 
\textit{since their time evolution is ruled by the Hamiltonian of the system}.
In the Heisenberg picture, operators evolve in time via Heisenberg 
equation, where the Hamiltonian is present. For an arbitrary operator $\hat O$, it reads
\begin{equation}
 i \frac{d \hat O}{d t} = [ \hat O, \hat H ] + i \frac{\partial}{\partial 
t} \hat O(t),  \label{Heis}
\end{equation}
where the last term accounts for possible explicit time dependence of the 
operator.

\subsection{Spectral representation}\label{sec:represp}

So far we have presented the Green's function in the time domain. But very 
often it is convenient to represent it in the energy 
domain. For example, when our system is at equilibrium or when the 
Hamiltonian is time-independent\footnote{If there is time translational symmetry, 
it is possible to describe the system via time differences $t-t'$ and perform a 
Fourier transform to represent the Green's function in 
the energy domain. Similarly, in the presence of spacial translational symmetry, 
the representation in the momentum space is also convenient.}. 
For such cases the Green's function will depend
only on time differences $t-t^\prime$ and we can perform a Fourier transform. 
To illustrate this, let us first consider the spectral representation in the 
special case of a free particle Hamiltonian, which can be written as
\begin{eqnarray}
H=\sum_m \varepsilon_{m} c_m^\dagger c_m, 
\end{eqnarray}
where $c_m^\dagger$ ($c_m$) creates (annihilates) and electron in the $m$-th 
single-particle eigenstate of the system with energy $\varepsilon_m$.

In the Heisenberg picture, using Eq.~\eqref{Heis}, the equations of motion 
of our operators are
{
\begin{align}
 \frac{d c_n}{dt}&=-i\left[c_n,H\right]=-i\varepsilon_n c_n\\
 \frac{d c^\dag_n}{dt}&=-i\left[c^\dag_n,H\right]=i\varepsilon_n c^\dag_n. 
\end{align}
Therefore, the creation and the annihilation operators evolve as $c_n(t)= e^{-i 
\varepsilon_{n} t} \,c_n$ and $c^\dagger_{n}(t)=e^{i \varepsilon_{n} t} \, 
c^\dagger_{n}  $. From these expressions, the retarded and advanced Green's 
functions of Eq.~\eqref{Gretard} and \eqref{Gadvance} for the free-particle 
case are simple functions of the time difference $t-t^\prime$:
\begin{align}
G_{nn'}^r(t-t')&=-i\,\theta(t-t')e^{-i\,\varepsilon_{n}(t-t')}\delta_{nn'}  
\label{Gretardnonint}\\
 G_{nn'}^a(t-t')&=\;\;\;i\,\theta(t'-t) e^{i\,\varepsilon_{n}(t'-t)}\delta_{nn'} 
 \label{Gadvancenonint} \,,
\end{align}
where we have used $\langle \{c_n, c^\dagger_{n'}\} \rangle =\delta_{nn'} $. 
Note that the Green's function is diagonal in the energy 
basis, which does not happen in the general interacting case, where the time 
evolution of the single particle operator involves different states. Here we 
assumed that the particle is in an eigenstate of a noninteracting Hamiltonian.

To write the spectral representations of \eqref{Gretardnonint} and 
\eqref{Gadvancenonint}, let us consider the integral representation 
of the Heaviside step function:
\begin{equation}
 \theta(t-t') = -\frac{1}{2\pi \,i} \int_{-\infty}^{\infty} d\omega 
\frac{e^{-i \omega (t-t')}}{\omega + i\eta}  \,, 
\end{equation}
where $\eta\to0^+$ is a positive infinitesimal real number. Inserting this 
expression in \eqref{Gretardnonint}, we obtain 
\begin{equation}
 G_{nn}^r(t - t')= \frac{1}{2\pi} \int_{-\infty}^{\infty} d\omega 
 \frac{ e^{-i (\omega + \varepsilon_n) (t-t')} }{\omega + i\eta} 
\,. 
\end{equation}
 
By performing a change of variables $\omega+\varepsilon_n\to\omega$, we have
\begin{equation}
 G_{nn}^r(t - t')= \frac{1}{2\pi} \int_{-\infty}^{\infty} d\omega 
 \frac{ e^{-i \omega (t-t')} }{\omega-\varepsilon_n + i\eta} 
 \,. \label{eq:Gnnt}
\end{equation}
Since $G_{nn}^r(t - t')$ is the Fourier transform\footnote{Here we define the Fourier 
transform of the retarded Green's function as %
\begin{align}
 G_{ij}^r(t-t')&=\frac{1}{2\pi}\int_{-\infty}^{\infty} d\omega \, 
e^{-i\omega (t-t')} G_{ij}^r(\omega)\\
G_{ij}^r(w)&=\int_{-\infty}^{\infty} dt \, e^{i\omega t} G_{ij}^r(t) \,. 
\label{Fourier}
\end{align}}  
 of $G^r(\omega)$, we can identify the latter in the integrand of Eq.~\eqref{eq:Gnnt},
\begin{equation}
 G_{nn}^r(\omega)= \frac{1}{\omega - \varepsilon_{n} + i\eta} \,.
\end{equation}
Analogously, we obtain for the noninteracting advanced Green's function,
\begin{equation}
 G_{nn}^a(\omega)= \frac{1}{\omega - \varepsilon_{n} - i\eta} \,.
\end{equation}

The Fourier transforms of the retarded/advanced Green's functions have different 
analyticity properties. This is a consequence of causality, expressed in the step 
functions of Eq.~\eqref{Gretard} and \eqref{Gadvance}. 
The retarded(advanced) Green's function is analytic in the upper(lower) 
half of the complex $\omega$ plane and has poles in the lower(upper) half plane, 
corresponding to the eigenenergies in this simplified example, and single-particle 
excitations in the more general case.

Converting to a site basis, $G_{ij}=\sum_{n} \langle i | n \rangle \langle n | j \rangle G_{nn}$, 
thus we obtain
\begin{equation}
 G_{ij}^{r/a}(\omega)= \displaystyle\sum_n \frac{\langle i |  n \rangle \langle 
n | j \rangle}{\omega - \varepsilon_{n} \pm i\eta}.
\end{equation}

There are many physical properties hidden in the Green's function. At this 
point we can extract at least two important properties of the retarded and 
advanced Greens functions:
\begin{enumerate}
 \item \textit{For the noninteracting Hamiltonian, the poles of the Green's 
function correspond exactly to the eigenenergies}. This can be immediately noticed 
since $\varepsilon_n$ was assumed to be the eigenenergy of the free 
particle system, governing the time evolution of the creation and annihilation 
operators. 
This property refers only to the simplified case of a noninteracting Hamiltonian. 

\item \textit{The imaginary part\footnote{One should have in mind that the imaginary
part of a matrix $A$ is $\textrm{Im}(A) = (A-A^\dagger)/(2i)$. We thank K. Pototzky 
for this remark.}
of the diagonal ($j=i$) retarded or 
advanced Green's function provides the local density of states of the system}:
\begin{center}
\begin{equation}
  \rho_i(\omega) = \mp \frac{1}{\pi} \textrm{Im} \{ G_{ii}^{r,a}(\omega)\}   \,. 
\label{eq:DOS} 
\end{equation}
\end{center}
Here we used the Cauchy relation.\footnote{Limits of improper integrals can be 
obtained by the principal value of the Cauchy relation
\begin{equation}
  \lim_{\eta\to0}\frac{1}{\omega-\varepsilon \pm i\eta} = 
\mathcal{P. V.}\left( \frac{1}{\omega-\varepsilon} \right) \mp i \pi 
\delta(\omega-\varepsilon) \,, 
\end{equation} 
due to the improper nature of the integrals of $G^{r/a}$, e.g. 
Eq.~\eqref{eq:Gnnt},
with poles in different halfplanes. 
The imaginary part of the diagonal retarded/advanced Green's function recovers
the local density of states of a discrete spectrum, 
$\rho_i(\omega) = \displaystyle\sum_n |\langle n| i \rangle|^2 
\delta(\omega-\varepsilon_n)$.}
\end{enumerate}

To generalize property \textit{1}, let us consider the expansion of the 
operators 
in the complete basis of a generic Hamiltonian. It is possible to show that the poles of the 
retarded/advanced Green's function contain information about the spectrum of 
the single-particle excitations (i.e., a single electron excitation) of the 
system. To show this, let $H$ be the Hamiltonian of the \textit{interacting 
many-body system}.  The Schr\"odinger equation is 
$H |n\rangle=\varepsilon_n | n\rangle$, where 
$|n\rangle$ and $\varepsilon_n$ are the many-body eigenstates and 
eigenenergies, respectively. Note that $|n\rangle$ forms a complete basis with closure relation 
\begin{eqnarray}\label{closure}
 \sum_n |n\rangle \langle n|=1.
\end{eqnarray}
Within the Heisenberg picture, a given operator $A(t)$ evolves from 
$t^\prime$ to $t$ as $A(t)=e^{iH(t-t^\prime)}A(t^\prime)e^{-iH(t-t^\prime)}$. If 
$H$ is time-independent, the evolution depends only on the difference 
$t-t^\prime$. The Green's function \eqref{Gretard} becomes

\begin{widetext}
\begin{eqnarray}\label{Gretard_int}
 G_{ij}^r(t,t')&=& -i \theta(t-t') \Big\langle \left\{ 
e^{iH(t-t^\prime)}c_i(t')e^{-iH(t-t^\prime)},  c_j^\dagger(t') \right\} 
\Big\rangle\nonumber\\
&=&-i 
\theta(t-t')\Big\langle\left[e^{iH(t-t^\prime)}c_i(t')e^{-iH(t-t^\prime)}
c_j^\dagger(t')+c_j^\dagger(t')e^ { iH(t-t^\prime)} 
c_i(t')e^{-iH(t-t^\prime)}   \right]\Big\rangle\nonumber\\
&=& -i 
\theta(t-t')\sum_{m}\Big\langle\left[e^{iH(t-t^\prime)}c_i(t')e^{-iH(t-t^\prime)
} |m\rangle \langle m | c_j^\dagger(t')+c_j^\dagger(t')e^ { 
iH(t-t^\prime)}|m\rangle \langle m | c_i(t')e^{-iH(t-t^\prime)}   
\right]\Big\rangle\nonumber\\
&=&-i\theta(t-t')\sum_{m}\Big\langle\left[e^{
-i\varepsilon_m(t-t^\prime)} e^{iH(t-t^\prime)}c_i(t') |m\rangle 
\langle m | c_j^\dagger(t')+e^ { 
i\varepsilon_m(t-t^\prime)}c_j^\dagger(t')|m\rangle \langle m | 
c_i(t')e^{-iH(t-t^\prime)}   
\right]\Big\rangle\nonumber\\
&=&-i\frac{1}{Z}\theta(t-t')\sum_{nm} e^{-\beta\varepsilon_n} \left[e^{
-i\varepsilon_m(t-t^\prime)} \langle n| e^{iH(t-t^\prime)}c_i(t') |m\rangle 
\langle m | c_j^\dagger(t')|n\rangle+ e^ { 
i\varepsilon_m(t-t^\prime)}\langle n|c_j^\dagger(t')|m\rangle \langle 
m | c_i(t')e^{-iH(t-t^\prime)}|n \rangle  \right]\nonumber\\
&=&-i\frac{1}{Z}\theta(t-t')\sum_{nm} e^{-\beta\varepsilon_n}\left[e^{
-i(\varepsilon_m-\varepsilon_n)(t-t^\prime)} \langle n|c_i(t') |m\rangle 
\langle m | c_j^\dagger(t')|n\rangle+ e^ { 
i(\varepsilon_m-\varepsilon_n)(t-t^\prime)}\langle n|c_j^\dagger(t')|m\rangle 
\langle m | c_i(t')|n \rangle   
\right].\nonumber\\
\end{eqnarray}
\end{widetext}
In the lines above we have performed the quantum statistical average $\langle 
A\rangle=Z^{-1}Tr [e^{-\beta H}A]$, where $Z$ is the partition function and 
$\beta$ is proportional to the inverse of the temperature.
For the diagonal Green's function $j=i$ we obtain,

\begin{widetext}
\begin{eqnarray}\label{Gretard_int1}
 G_{ii}^r(t,t')&=& -i\frac{1}{Z}\theta(t-t')\sum_{nm} 
e^{-\beta\varepsilon_n}\left[e^{ -i(\varepsilon_m-\varepsilon_n)(t-t^\prime)} 
|\langle n|c_i(t') |m\rangle|^2+ e^ { 
i(\varepsilon_m-\varepsilon_n)(t-t^\prime)}|\langle m|c_j^\dagger(t')|n\rangle 
|^2 \right]\nonumber\\
&=&-i\frac{1}{Z}\theta(t-t')\sum_{nm} \left[e^{-\beta\varepsilon_n}e^{ 
-i(\varepsilon_m-\varepsilon_n)(t-t^\prime)} |\langle n|c_i(t') |m\rangle|^2+ 
e^{-\beta\varepsilon_m}e^ { i(\varepsilon_n-\varepsilon_m)(t-t^\prime)}|\langle 
n|c_j^\dagger(t')|m\rangle |^2 \right]\nonumber\\
&=&-i\frac{1}{Z}\theta(t-t')\sum_{nm} |\langle n|c_i(t') |m\rangle|^2e^{ 
-i(\varepsilon_m-\varepsilon_n)(t-t^\prime)} \left(e^{-\beta\varepsilon_n}
+ e^{-\beta\varepsilon_m}\right)\nonumber\\
\end{eqnarray}
\end{widetext}
We can now set $t^\prime=0$ and take the Fourier transform, as we did for the 
noninteracting case:
\begin{eqnarray}\label{Gretard_int_w}
 G_{ii}^r(\omega)&=&\frac{1}{Z}\sum_{nm} \frac{|\langle n|c_i(0) 
|m\rangle|^2}{\omega-(\varepsilon_m-\varepsilon_n)+i\eta}\left(e^{
-\beta\varepsilon_n} + e^{-\beta\varepsilon_m}\right).\nonumber\\
\end{eqnarray}

This expression is known as the Lehmann or spectral representation of the 
Green's functions \cite{Bruus}. Following property number \textit{2} of the 
retarded/advanced Green's functions shown above, from the diagonal Green's 
function we can calculate the local density of states:
\begin{eqnarray}\label{rho_int_w}
 \rho_i(\omega)&=&\frac{1}{Z}\sum_{nm} |\langle n|c_i(0) 
|m\rangle|^2\left(e^{
-\beta\varepsilon_n} + 
e^{-\beta\varepsilon_m}
\right)\nonumber\\
&&\times\delta[\omega-(\varepsilon_m-\varepsilon_n)].
\end{eqnarray}

It is possible to show that Eq.~\eqref{eq:DOS} is recovered when 
considering a noninteracting Hamiltonian. In this case the Hamiltonian is separable, 
and the many-particle eigenstates are a antisymmetrized product of single-particle states.
The expectation value in \eqref{rho_int_w} will connect states $m$ that have one 
additional electron in the site $i$ compared to state $n$, thus $E_m = E_n + \varepsilon_i$, 
where $\varepsilon_i$ is the energy of an additional bare electron at site $i$. 
Careful manipulation of  \eqref{rho_int_w} and the partition function $Z$ results in
a local density of states independent of the temperature, with poles at 
single-particle energies $\varepsilon_i$.\\

Among the many interesting properties of the \textit{interacting} Green's function 
\eqref{Gretard_int1} we can also emphasize that:
\begin{enumerate}
 \item The poles of the \textit{interacting} Green's function are exactly 
 at the many-body excitations $\varepsilon_m-\varepsilon_n$ of the system;
\item In contrast with the noninteracting case, both the Green's function 
\eqref{Gretard_int1}  and the local density of states depend on the 
temperature. This is characteristic of interacting systems.
\end{enumerate}

Although we have presented a more robust formalism, in the examples treated 
in this article, we will deal only with noninteracting Hamiltonians, 
neglecting Coulomb interactions, and our local density of states will map the 
spectra of each Hamiltonian.
\section{The equation of motion technique}
 
One way of obtaining the Green's function is to determine its time evolution 
via equation of motion (EOM) technique. Using the Heaviside function $\theta(t-t')$ 
and the Heisenberg equation of motion for the operator $c_i(t)$, we derive the 
retarded Green's function \eqref{Gretard} with respect to time:

\begin{eqnarray}
 i \partial_t G^r_{ij}(t,t') &=& i (-i) \partial_t \theta(t-t') 
\big\langle \big\{ c_i(t), c_j^\dagger(t') \big\} \big\rangle \nonumber \\
&& -i \theta(t-t') \big\langle \{ i \dot c_i(t) , c_j^\dagger(t') \} 
\big\rangle \nonumber \\
 %
&= &\delta(t-t')\delta_{ij} \nonumber \\ 
&& -i \theta(t-t') \langle \{ \left[c_i, H \right](t), c_j^\dagger(t') \} 
 \rangle \label{EOMr} \,.
\end{eqnarray}

In the last line, on the right-hand side (rhs) of Eq.~\eqref{EOMr}, there 
is one propagator that yet needs to be determined, which depends on the 
commutator of the operator $c_i$ with the Hamiltonian. We first note that this 
result is not restricted to $G^r$ but rather, is general:
the equation of motion will couple the original Green's function to a new one.
In addition, its dependence with the Hamiltonian will influence the dynamics.

From now on, we shall use more frequently the spectral representation for 
the Green's functions. Therefore, we present a simplified notation for 
the retarded Green's function in the energy domain, 
adapted from Zubarev \cite{Zubarev},
\begin{equation}
    G^r_{ij}(\omega)= \langle\langle \, c_i;c_j^\dagger \, \rangle\rangle 
\,.
\end{equation}

Performing the Fourier transform defined in Eq.~\eqref{Fourier} on 
Eq.~\eqref{EOMr}, we will obtain an factor $i\omega$ on the left coming 
from the time derivative. Since the Fourier transform of the 
$\delta$-function is the unity,\footnote{
\begin{equation}
 \delta(t-t')=\frac{1}{2\pi} \int_{-\infty}^{\infty} d\omega \, 
e^{-i\omega (t-t')} \qquad \mbox{and} \qquad 
 1= \int_{-\infty}^{\infty} dt \, e^{i\omega t} \delta(t) \,.
\end{equation}}
the spectral representation of the equation of motion \eqref{EOMr} acquires the form 
\begin{equation}
    \omega G^r_{ij}(\omega)= \delta_{ij} + \langle\langle [c_i,H] ; c_j^\dagger 
\rangle\rangle \,.
    \label{EOMrw}
\end{equation}

We stress that the presence of the commutator on the rhs of Eqs. \eqref{EOMr} 
and \eqref{EOMrw} tells us that \textit{the dynamics of the 
Green's function is fully determined by the Hamiltonian of the system}.

\subsection{Simple example: the non-interacting linear chain}

Let us consider a linear chain described by the non-interacting 
Hamiltonian containing a single orbital (energy) per site and a  kinetic 
term that connects all nearest-neighbor sites via a hopping 
parameter $t$
\begin{eqnarray}
 H &=& \sum_{l} \varepsilon_{l} c^{\dagger}_l c_l +  \sum_{l} ( t_{l+1,l} 
c^{\dagger}_l c_{l+1} + t_{l,l+1} c^{\dagger}_{l+1} c_l ) \nonumber \\
 &=& h_{pot} + h_{cin} . \label{Ham_potcin}
\end{eqnarray}

The first sum in Eq.~\eqref{Ham_potcin} corresponds to a local external 
potential that is diagonal in a base of sites. The second term corresponds to the 
kinetic energy, describing the destruction of a particle in the site $l+1$ and 
creation of another particle in the site $l$ with probability amplitude 
$t_{l+1,l}$. The third term describes the reverse process. The Hamiltonian is 
hermitian as it represents an observable, namely, the total energy of the system. 
To assure hermicity, $t^*_{l,l+1}=t_{l+1,l}$.

To calculate the commutator $[c_i,H]$ we simply use commutation 
rules\footnote{One may find useful to apply $[AB,C]=A\{B,C\}-\{A,C\}B$ and $[A,B]=-[B,A]$.} 
listed in Sec.~\ref{sec:represp}, from which we obtain
\begin{align}
[c_i,h_{pot}]=&\sum_{l} \varepsilon_{l} [ c_i, c^{\dagger}_{l} c_{l} ] = 
\sum_{l} \varepsilon_{l} \delta_{i l} c_{l} = \varepsilon_{i} c_{i} \,,\\
[c_i,h_{cin}]=& \sum_{l}\left\{ t_{l+1,l}  [ c_i, c^{\dagger}_{l} 
c_{l+1}]+t_{l,l+1}  [ c_i, c^{\dagger}_{l+1} c_{l}]\right\} \nonumber \\ 
=& \sum_{l}\left( t_{l+1,l} \delta_{i,l} c_{l+1}+t_{l,l+1} \delta_{i,l+1}  
c_{l}\right) \nonumber \\ 
=& \sum_{l}\left(t_{i+1,i}c_{i+1}+t_{i-1,i}c_{i-1}\right) \nonumber \\ 
=& \sum_{j=\pm 1} t_{i+j,i} c_{i+j}.
\end{align}

We now introduce these commutators into the equations of motion (EOMs)
\eqref{EOMr} or \eqref{EOMrw}. In the energy domain\footnote{In the time 
domain the EOM  has the form 
\begin{equation}
  (i \partial_t - \varepsilon_{i}) G^r_{ij}(t - t') = 
\delta(t-t')\delta_{ij} + \sum_{k=\pm 1} t_{i+k,i} G^r_{i+k,j}(t - t') 
\,.
\end{equation}
}, see Eq.~\eqref{EOMrw}, we have

\begin{equation}
 (\omega- \varepsilon_{i} + i \eta)\, G_{ij}^r(\omega) = \delta_{ij} + 
\sum_{k=\pm 1} t_{i+k,i} G^r_{i+k,j}(\omega) \,, \label{EOMcin}
\end{equation}
where the propagator $G^r_{ij}(\omega)$ couples to other 
propagators through  first neighbor hopping. In this work we 
will consider only Hamiltonians that couple \textit{nearest 
neighbors} in different geometries. As the reader becomes familiar
with  the technique, its operation and usage become clearer.

It is important to emphasize that the local potential and the kinetic 
energy are single particle operators and do not produce many-particle 
Green's functions.  In a more general case where the Hamiltonian 
has two-particle operators, \textit{i.e.}, a product of four operators, it 
will generate multi-particle Green's functions. The resulting system of 
coupled Green's functions is a priori, infinite, but for practical 
purposes it is truncated at some level. Despite their importance in 
condensed matter physics, many-particle Hamiltonians are outside the scope 
of this work, but can be found elsewhere, \textit{e.g.} 
Refs.\cite{Bruus} and \cite{HaugJauho} and references therein. 
In the example treated here, Hamiltonians are noninteracting and we 
can find exact  solutions (at least numerically) for the Green's 
functions. Even for noninteracting systems, few examples grant an analytical 
expression for the Green's function. For the others we can at least obtain exact 
numerical solutions. Indeed, numerical solutions are the  main motivation of 
this work. 

\subsection{Two-site chain: the hydrogen molecule}

The simplest finite lattice has only two sites, see Fig.~\ref{fig:2sites}(a).
Before deriving an exact expression for the Green's functions of this 
system, let us review its relevance in quantum chemistry as a prototype 
of the molecular bond between two hydrogen nuclei.
In this model, each atom has its $s$-type orbital localized around 
its H nucleus with energy $\varepsilon_0$, shown in Fig.~\ref{fig:2sites}(b). 
The proximity of the two atoms allows for the hybridization of their individual 
orbitals with overlap matrix element (hopping) $t$. This coupled system 
has two solutions, two molecular orbitals with even and odd symmetry with 
respect to spatial inversion,\footnote{We should notice that we fully neglect spin-orbit 
contributions in the Hamiltonian. Thus in this problem spatial degrees 
of freedom are decoupled from spin, since nor the kinetic energy nor the local potential couples to 
the spin of the particles.} known as bonding and anti-bonding states. 
They have energies $\varepsilon_0\mp |t|$, illustrated in the energy diagram 
of Fig.~\ref{fig:2sites}(c).
\begin{center}
\begin{figure}[ht]
 \centering\includegraphics[width=\columnwidth]{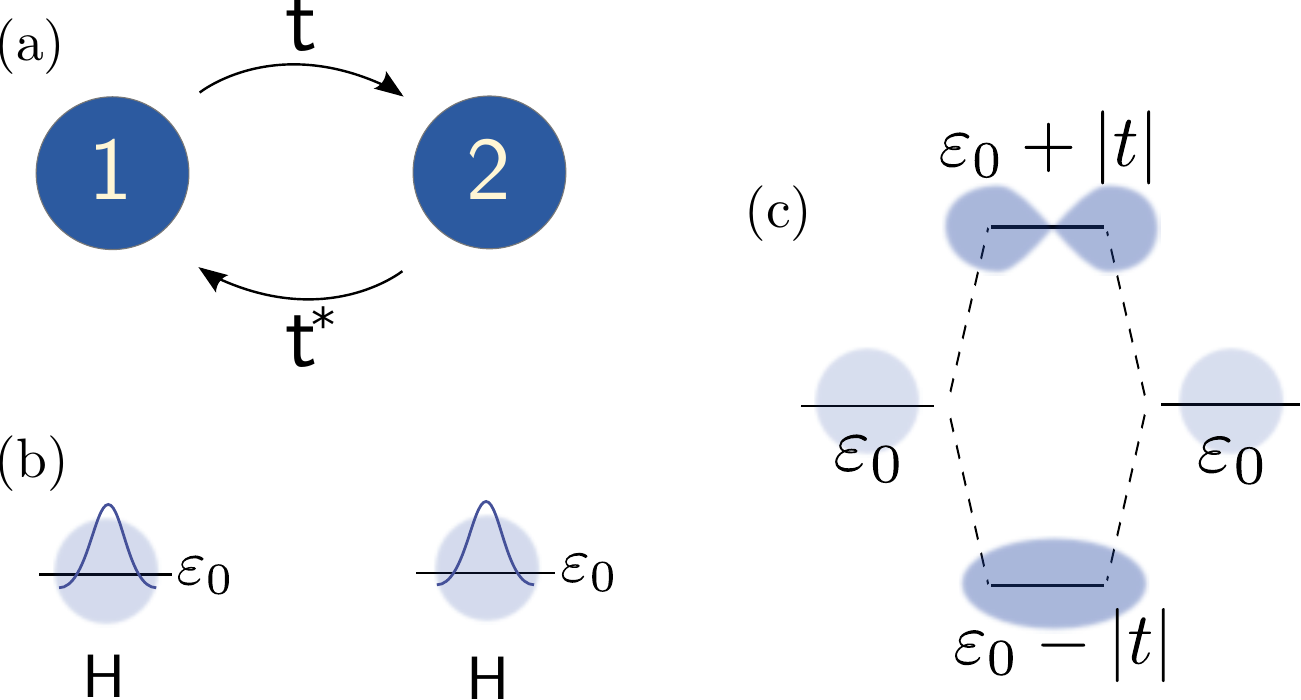}
 \caption{(a) Finite chain with two sites and overlap matrix elements 
 $t$ e $t^*$. (b) The two-site system is a prototype model in chemistry, 
 where each site is pictured as a hydrogen nucleus with a single $s$-orbital localized 
 around it with energy $\varepsilon_0$. (c) Energy level diagram, where we see the 
 formation of two molecular orbitals. The presence of hybridization generates
 an even ground-state known as the bonding state, and the excited anti-bonding state, 
 which has a node in the spatial wavefunction. Figure adapted from 
Ref.~\cite{Cuevasbook}.}
 \label{fig:2sites}
\end{figure}
\end{center}

For the present case, with $N=2$, the Hamiltonian \eqref{Ham_potcin} reads
\begin{equation}
 H =\varepsilon_0 \left( n_1 + n_2 \right) + t c^{\dagger}_{2} c_{1} +
 t^* c_{1}^{\dagger}c_{2} \,,
 \label{H2}
\end{equation}
where we define the local energy $\varepsilon_0$, the number operator 
$n_i=c^\dagger_i c_i$ and the hopping matrix element $t_{21}=t$. 
In this problem, we can distinguish the Hamiltonian for the two isolated sites, 
$\mathbf{h}_0$, and a perturbation 
(inter-site coupling) $\mathbf{V}$. This perturbative perspective allows 
us to write a \textit{Dyson equation} for the Green's function of the 
system, as we will develop below. The matrix representing the 
Hamiltonian \eqref{H2} on the local orbitals basis $\{ 1,2\}$ acquires the 
form
\begin{eqnarray}
    \mathbf{H}&=&\mathbf{h}_0+\mathbf{V}=
    \begin{pmatrix}
    \varepsilon_0 & 0 \\
    0 & \varepsilon_0
    \end{pmatrix}
+    
    \begin{pmatrix}
    0 & t^* \\
    t & 0
    \end{pmatrix} \,.\label{Mform}
\end{eqnarray}

The energies of the molecular orbitals $\varepsilon_0\mp |t|$ are easily 
obtained by diagonalizing the Hamiltonian above.

Returning to the explicit calculation of the Green's functions, we see 
that the local Green's function for the first site, 
$G_{11}^r(t,t')=-i\theta(t-t')\langle\{c_1(t),c_1^\dagger(t')\} \rangle$ 
is coupled to the non-local Green's function (propagator) 
$G_{21}^r(t,t')=-i\theta(t-t')\langle\{c_2(t),c_1^\dagger(t')\} \rangle$, 
introduced by the commutators indicated in Eqs.~\eqref{EOMr} and 
\eqref{EOMcin}. In time domain we obtain the following equations of motion (EOMs),
\begin{align}
 (i\partial_t - \varepsilon_0) G^r_{11}(t,t') &= \delta(t-t') + 
t\,G^r_{21}(t,t') \,,\\
 (i\partial_t - \varepsilon_0) G^r_{21}(t,t') &= t^*\,G^r_{11}(t,t') \,,
\end{align}
while in energy domain we have, 
\begin{align}
 (\omega - \varepsilon_0 + i\eta)G^r_{11}(\omega)&=1+t\,G^r_{21}(\omega) \label{eq:H2G11}\\ 
 (\omega - \varepsilon_0 + i \eta)G^r_{21}(\omega)&=t^*\,G^r_{11}(\omega) \,. \label{eq:H2G12}
\end{align}

From the equations above we see that is useful to introduce the undressed 
local Green's functions for the isolated sites  (that can be obtained by 
setting $t=0$ in the equations above), 
\begin{equation}
 g^r(\omega) = \frac{1}{\omega - \varepsilon_0 + i \eta} = g_1^r(\omega) = 
g_2^r(\omega) \,, \label{glivre} 
\end{equation}
where we define the lowercase $g$ referring to the Green's 
function of an isolated site. This function, which we name 
\textit{undressed} Green's function, is diagonal on the isolated site 
basis, similarly to the unperturbed Hamiltonian. For the hydrogen 
molecule [Fig.~\ref{fig:2sites}(a)], the  \textit{dressed} Green's 
function  exhibits non-diagonal terms due to the couplings. In
matrix form, the \textit{undressed} and \textit{dressed}  Green's 
functions read
\begin{align}
    \mathbf{g}^r=
    \begin{pmatrix}
    g_1^r & 0 \\
    0 & g_2^r
    \end{pmatrix} \qquad \mbox{and} \qquad
    \mathbf{G}^r=
    \begin{pmatrix}
    G_{11}^r & G_{12}^r \\
    G_{21}^r & G_{22}^r
    \end{pmatrix}\,,
    \label{eq:gGmatrix}
\end{align}
where by inversion symmetry around the center of the mass of the 
molecule, we can write $G^r_{22}(\omega)=G^r_{11}(\omega)$.

In terms of the undressed Green's function \eqref{glivre}, we obtain the 
coupled system of equations
\begin{align}
 G_{11}(\omega)&=g^r(\omega)+g^r(\omega)\,t\,G^r_{21}(\omega) 
\label{G11H2} \\
 G_{21}(\omega)&=g^r(\omega)\,t^*\,G^r_{11}(\omega) \label{G21H2}\,.
\end{align}

These linear equations are rewritten more compactly in a matrix notation, 
\textit{i.e.}, in terms of Eq.~\eqref{eq:gGmatrix},
\begin{equation}
 \mathbf{G}^r=  \mathbf{g}^r + \mathbf{g}^r \, \mathbf{V} \, \mathbf{G}^r, \label{eq:Grdyson}
\end{equation}
where the coupling potential $\mathbf{V}$ was defined in Eq.~\eqref{Mform}. 
In this form, the \textit{dressed} Green's function $\mathbf{G}^r$, is obtained 
by isolating it as
\begin{equation}
 \mathbf{G}^r = \left(1 - \mathbf{g}^r \, \mathbf{V} \right)^{-1} \, \mathbf{g}^r \, .
\end{equation}

To find the explicit expression for the local site Green's function we 
can eliminate the non-diagonal propagator by replacing Eq.~\eqref{G21H2} into 
Eq.~\eqref{G11H2}, or equivalently, \eqref{eq:H2G12} in \eqref{eq:H2G11}
\begin{equation}
 G_{11}(\omega) = \frac{g^r(\omega)}{ 1-|t|^2\,[g^r(\omega)]^2} =  
\frac{1}{\omega-\varepsilon_0-|t|^2\,g^r(\omega)+i\eta}\,. 
\label{eq:H2G11closed}
\end{equation}

In the last term of \eqref{eq:H2G11closed}, $g^r(\omega)$ can 
contribute with a real and a imaginary part in the denominator. 
This means that there can be a change of the position of the resonance 
energy $\varepsilon_0$ and a broadening of the correspondent peak. 
Since $g^r(\omega)$ is the function of an isolated site, its imaginary part is just
a $\delta$-like function, resulting in no effective broadening. 
In Fig.~\ref{gr:dos2} we plot the density of states, which is proportional 
to $\textrm{Im}[G_{11}^r]$ via Eq.\eqref{eq:DOS}. The broadening of the peaks 
was artificially increased with $\eta=0.01$ for visualization. Thus the 
final effect of the tunneling between the two sites on site $1$ 
is a change of the local energy $\varepsilon_0$ to $\varepsilon_0 \pm |t|$.
More generally, the coupling of a site to another structure causes a 
shift of the resonance to a new energy $\tilde\varepsilon_0$ a 
broadening $\Gamma$, i.e., 
$G_{11}(\omega) = \left(\omega - \tilde\varepsilon_0 + i \Gamma \right)^{-1}$.

In addition, Eq.~\eqref{eq:H2G11closed} can be rewritten as a sum of partial fractions,
\begin{equation}
 G_{11}(\omega) = \frac{1/2}{\omega - (\varepsilon_0 - |t|) + i \eta} + \frac{1/2}{\omega - (\varepsilon_0 + |t|) + i \eta} \,,
\end{equation}
where we identify the two eigenvalues of the molecule, shown in Fig.~\ref{fig:2sites}(c).
As discussed in Sec.~\ref{sec:represp}, the poles of the noninteracting 
Green's function correspond exactly to the eigenenergies, and the imaginary part 
leads to the density of states, shown in Fig.~\ref{gr:dos2}.

\begin{center}
\begin{figure}[ht]
 \centering\includegraphics[width=\columnwidth]{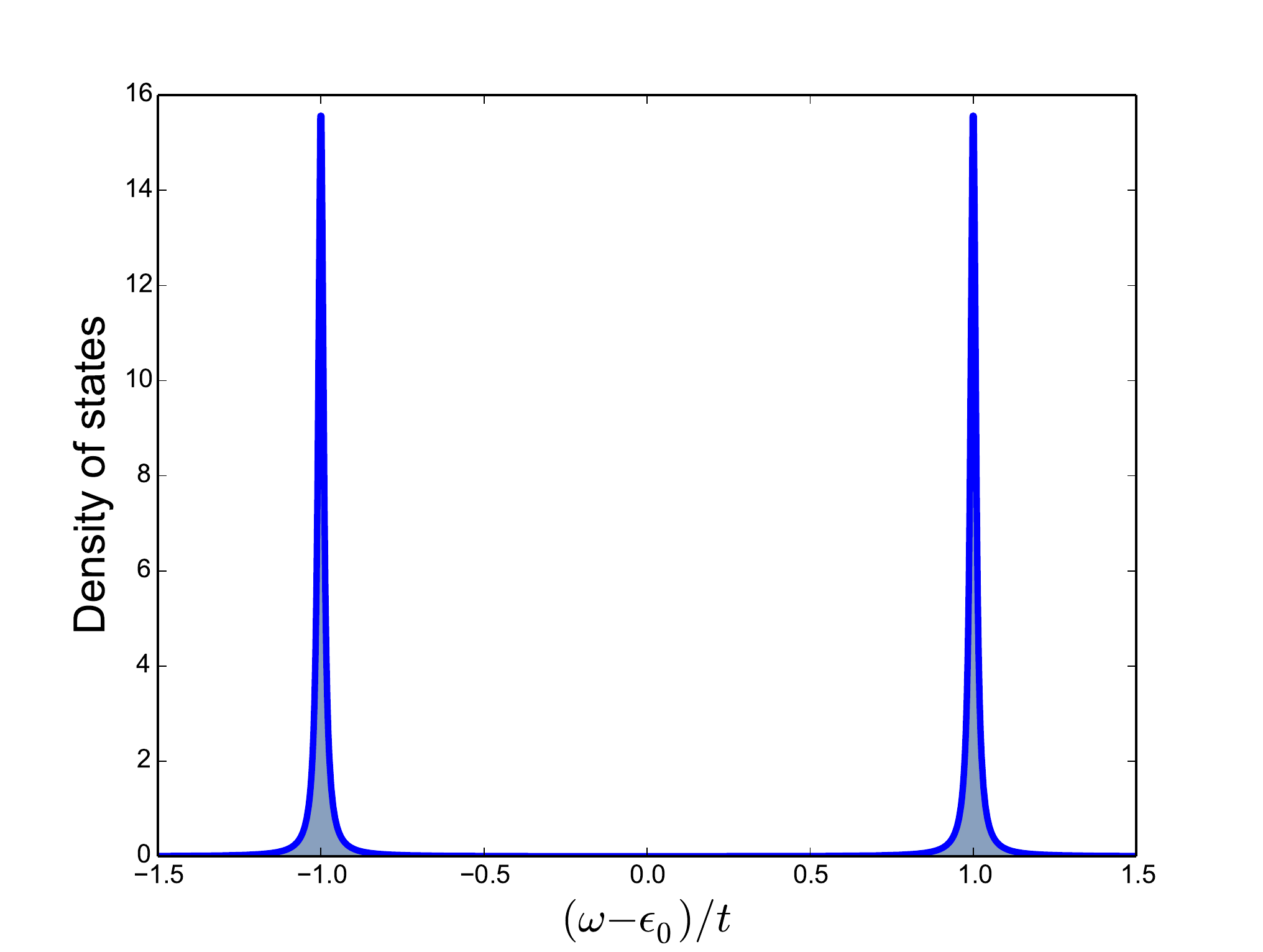}
 \caption{Density of states of the first site in the hydrogen molecule. 
 We have set a large $\eta=0.01$ for visualization of the broadening of the peaks. } 
 \label{gr:dos2}
\end{figure}
\end{center}

It is important to mention that, within the perturbative approach, 
the Green's function of the system can be obtained by a recursive relation 
called \textit{Dyson equation}:

\begin{equation}
  \mathbf{G}(\omega) = \mathbf{g}(\omega) +  \mathbf{g}(\omega)  \mathbf{\Sigma}(\omega)  \mathbf{G}(\omega) \,. \label{eq:Dyson}
\end{equation}
where $\mathbf{G}$ and $\mathbf{g}$ are the \textit{dressed} and 
\textit{undressed} (or \textit{bare}) Green's functions. In writing \eqref{eq:Dyson} we assumed that our problem 
allows a perturbative approach and that we can encapsulate the irreducible 
diagrams due to many-particle interactions in a operator called 
\textit{self-energy} $\mathbf{\Sigma}(\omega)$. The \textit{self-energy} is an 
energy-dependent operator that accounts for the effects of self-consistent 
interactions, the dynamic \textit{i.e.}, energy-dependent, renormalization of the 
single-particle states. This renormalization will change the position 
of the level, and its width. This broadening is frequently related with the 
inverse of the lifetime of the \textit{dressed} particle, the quasiparticle. 
For interacting problems and more complex structures, the determination of a 
consistent self-energy is a challenging problem \cite{Mahan, NEGFStef}.
In our example, see Eq.~\eqref{eq:Grdyson}, $\mathbf{V}$ has a simple structure and the coupling $t$ is 
a constant, thus interactions and additional complications in the Hamiltonian
are not yet present.

In the next examples we will practice the equations of motion analytically and 
later numerically, for extended linear lattices.

\subsection{Semi-infinite linear chain}

An interesting example that provides an analytical closed solution of the 
equations of motion is the semi-infinite linear chain, shown in 
Fig.~\ref{fig:semiinfin}. This extended lattice can be considered a simple model 
of a crystalline solid or a semi-infinite electrode in a junction. 
\begin{center}
\begin{figure}[ht] \centering
 \includegraphics[width=7cm]{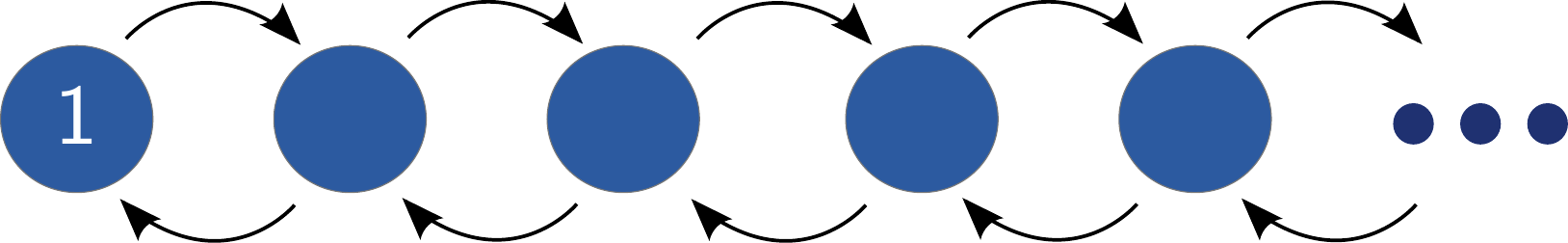}
 \caption{One-dimensional semi-infinite chain of atomic sites.}
 \label{fig:semiinfin}
\end{figure}
\end{center}

Note that the infinite number of sites prohibits direct diagonalization of the 
Hamiltonian or the resolvent operator, and the application of Eq.~\eqref{EOMcin} 
leads to an infinite hierarchy of propagators, with an infinite continued fraction 
structure. Already from early days of computational physics recursive techniques 
in tight-binding lattices were recognized as an efficient tool for the study 
of solids \cite{Haydock1980}. For instance, the workhorse in quantum transport,
the ``surface Green's function'' method approached in Sec.~\ref{sec:RGFs}, 
plays an essential role in the simulation of dynamic properties of materials. 

The decimation technique is a very useful tool for the recursive 
procedure. Basically, it is a strategy to approximate the solution of an 
infinite system starting from a finite one. 
This technique relies on finding a change of variables
that will bring your coupled equations of motion in the same form of a 
well known result. For instance, suppose we could add many sites to the 
hydrogen molecule, always renormalizing the Green's functions in a way 
to recover an effective site $\tilde 2$. Then one would have  
an effective hydrogen-like molecule, as illustrated in Fig.~\ref{fig:halfinf0} 
(note that the isolated sites are not identical). 
Here we assumed that we have already encapsulated a large number of 
sites into this 
effective site $\tilde 2$. In the asymptotic limit, this effective site 
gives the same answer of a semi-infinite lattice.
\begin{center}
\begin{figure}[ht] \centering
 \includegraphics[width=2.5cm]{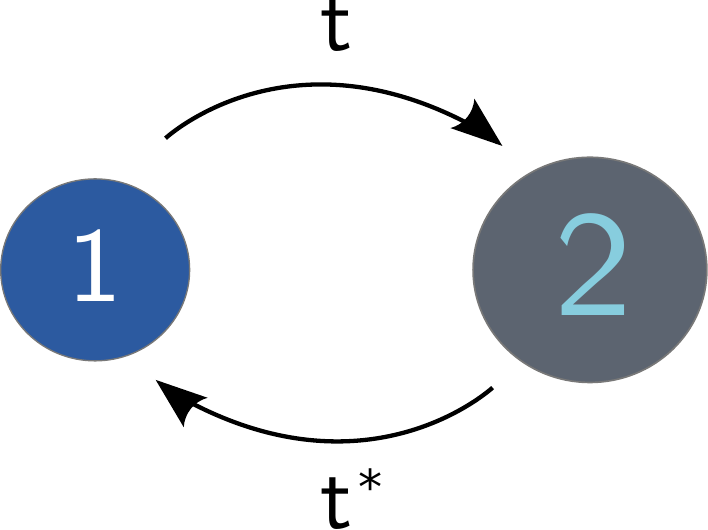}
 \caption{Effective hydrogen molecule to evaluate the Green's function of the 
 semi-infinite chain.}
 \label{fig:halfinf0}
\end{figure}
\end{center}

Let us then consider the effective two-site model, where 
one undressed surface site is coupled to an effective one. We have already 
developed the equations of motion of the two-site system, Eq.~\eqref{G11H2} 
and \eqref{G21H2}. For simplicity we will drop the frequency dependence and 
the retarded index in our notation. The equations of the effective two-site chain read
\begin{align}
G_{11} &= g_1 + g_1\, t \,G_{21} \label{eq:syssemiinf1} \\ 
G_{21} &= \tilde{G_2}\, t^* \,G_{11} \,, \label{eq:syssemiinf2}
\end{align}
where $g_1$ and $\tilde{G_2}$ are the undressed and the dressed effective Green's 
function.

In the limit of a infinite number of sites in the effective site $\tilde 2$,
the effective propagator $\tilde{G_2}$ describes itself the semi-infinite chain, 
i.e., $\tilde{G_2}=G_{11}$. With this observation, we solve the system in 
Eq.~\eqref{eq:syssemiinf1} and \eqref{eq:syssemiinf2}, finding a second-order 
equation for $G_{11}$:
\begin{equation}
 g_1 |t|^2 G_{11}^2 - G_{11} + g_1 = 0 \,. \label{eq:G11baskara}
\end{equation}

The two retarded solutions of Eq.~\eqref{eq:G11baskara} are given by
\begin{equation}
G_{11} = \frac{1}{2\,g_1 |t|^2} \left( 1 \pm \sqrt{1-4\,|t|^2 g_1^2} \right) \,,
\end{equation}
or, replacing the undressed function, Eq.~\eqref{glivre},
\begin{equation}
 G_{11}= \frac{\omega - \epsilon_0  + i\eta}{|t|^2} \left[1 \pm 
\sqrt{1-\frac{4|t|^2}{\left(\omega - \epsilon_0 + i\eta\right)^2}} \right]\,. 
\label{eq:semiG11w}
\end{equation}

We can determine the physical solution examining the analyticity properties of the 
Green's function \cite{Haydock1980}. In the asymptotic limit of $|\omega|\to\infty$ 
we must have a vanishing solution, therefore we choose
\begin{align}
 G_{11} &= \frac{\omega - \epsilon_0  + i\eta}{|t|^2} \left[1 - \sqrt{1-\frac{4|t|^2}{\left(\omega - \epsilon_0 + i\eta\right)^2}}\, \right].
\end{align}

One can verify that $G_{11}$ decays as $1/\omega$ in the asymptotic limit. 
Since the real and imaginary parts of the Green's functions are related by a Hilbert 
transform\footnote{The Hilbert transform is an improper integral, defined by the 
principal value 
\begin{equation}
 g(y) = \frac{1}{\pi} \mathcal{P. V.} \int_{-\infty}^{\infty} \frac{f(x) dx}{x-y} \,.
\end{equation}
For an analytic function in the upper plane, the Hilbert transform describes the 
relationship between the real part and the imaginary part of the boundary values.
This means that these functions are conjugate pairs.
Given a real-valued function $f(x)$, the Hilbert transform finds a imaginary part, 
a companion function $g(x)$, so that $F = f(x) + ig(x)$ can be analytically extended
to the upper half of the complex plane.}, this decay assures a bounded density of states \cite{Cuevasbook}. 
Note that, by factoring out $-1$ from the square root of \eqref{eq:semiG11w} we 
obtain the imaginary contribution, which is non-zero only in the region 
$|\omega - \epsilon_0| < 2|t|$, i.e., within the bandwidth. This gives the density 
of states of the edge, or ``surface'' site:
\begin{eqnarray}
 \rho_1(\omega) &\!=\!& - \frac{1}{\pi} {\rm Im } G_{11}(\omega)  \nonumber\\
 &\!=\!& \frac{1}{\pi |t|}\sqrt{1-\left(\frac{\omega-\varepsilon_0}{2|t|}\right)^2} \theta(2|t|-|\omega-\varepsilon_0|), \label{eq:dossemi} 
\end{eqnarray}
which forms a semi-circle, as illustrated in Fig.~\ref{fig:semiReIm}. In this graph we 
plotted $-|t|Im[G_{11}^r]$ to scale with the real part.

\begin{center}
\begin{figure}[ht]
 \centering\includegraphics[width=\columnwidth]{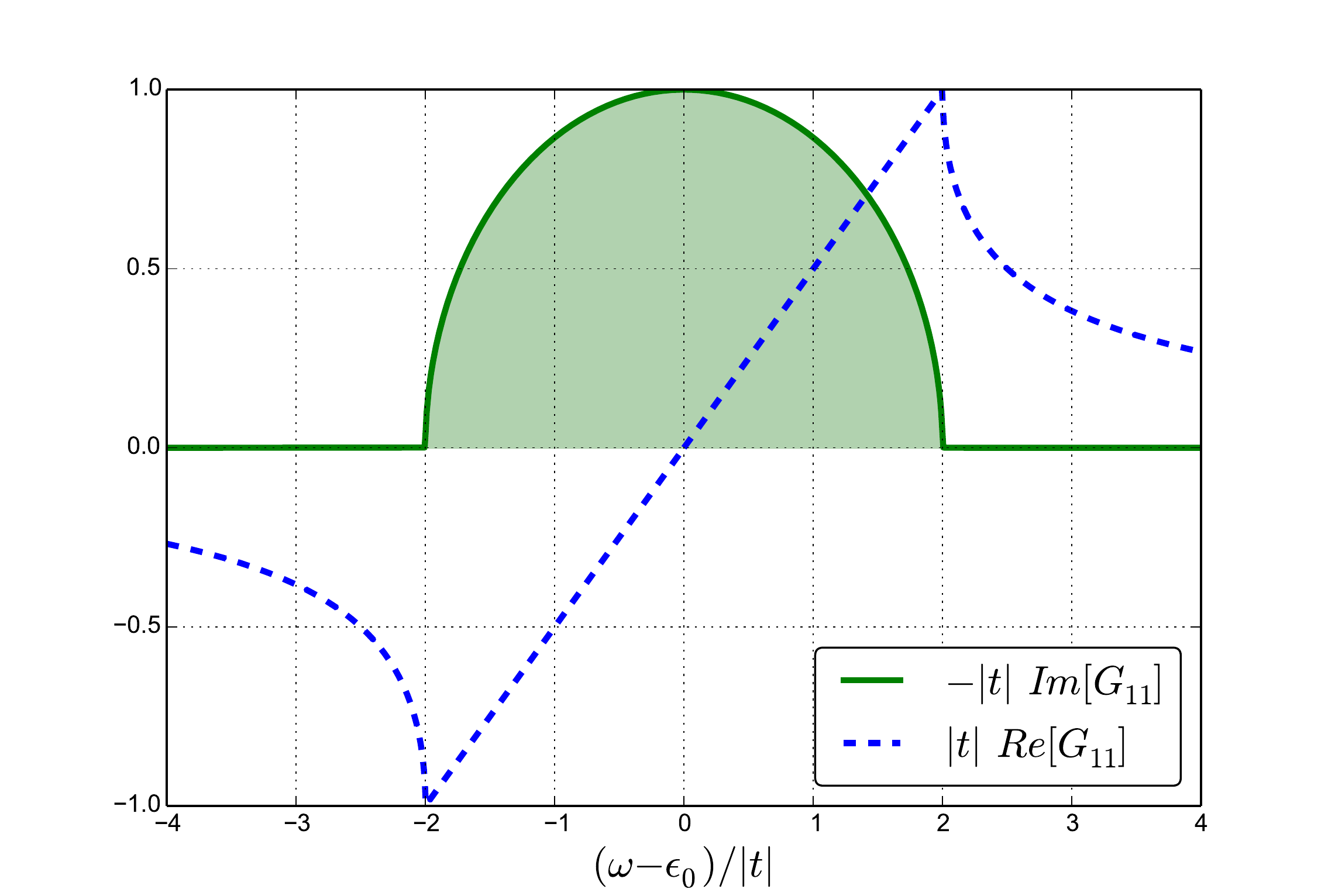}
 \caption{Real and imaginary parts of the surface Green's function of a linear chain.
 The imaginary part relates to the density of states, which is a semicircle bounded 
 by the bandwidth $2|t|$. In this example, $\eta=0.0001$.}
 \label{fig:semiReIm}
\end{figure}
\end{center}

\subsection{Infinite linear chain} \label{sec:infinite}

Another interesting model that allows analytical solution is the infinite 
linear chain. The band structure and density of states can be easily 
obtained in the tight-binding framework by considering Bloch  eigenfunctions 
\cite{Cini}. Here we will show how to obtain the DOS from the equations of 
motion. 

The infinite chain can be viewed as the coupling between two semi-infinite 
chains, as shown in Fig.~\ref{fig:inf}(a). This would correspond to two 
effective sites in a two-site model, as in Fig.~\ref{fig:inf}(b), with solution
\begin{equation}
 G_{11}=\frac{\tilde G_1}{1-\tilde G_1^2 |t|^2} \,, \label{eq:Ginf2}
\end{equation}
where $G_{11}$ is the diagonal dressed Green's function of the infinite 
lattice, while the effective propagator $\tilde G_1=\tilde G_2$ is the previous 
semi-infinite answer, Eq.~\eqref{eq:semiG11w}. 

\begin{center}
\begin{figure}[ht]
 \centering\includegraphics[width=\columnwidth]{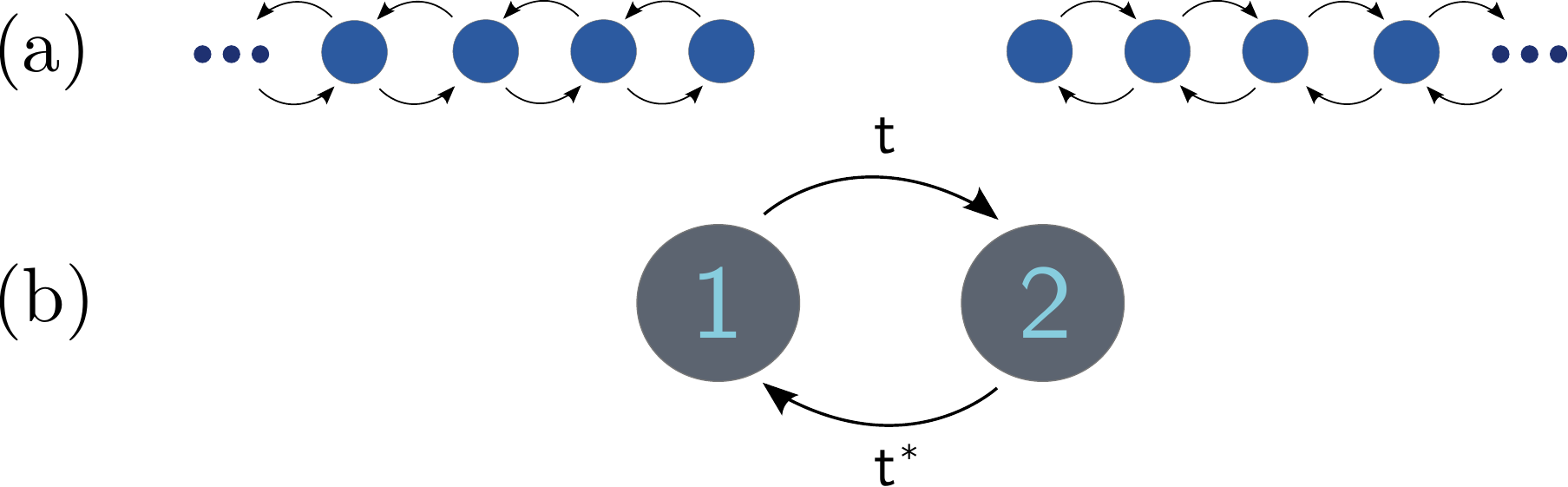}
 \caption{(a) Infinite linear chain pictured as the coupling of two semi-infinite 
 lattices.(b) Effective sites that encapsulate the semi-infinite chains.}
 \label{fig:inf}
\end{figure}
\end{center}

One might wonder if this solution is unique. Other couplings are possible, 
for example, in Fig.~\ref{fig:inf3}(a) we couple one undressed site with two 
semi-infinite lattices.

\begin{center}
\begin{figure}[ht]
 \centering\includegraphics[width=\columnwidth]{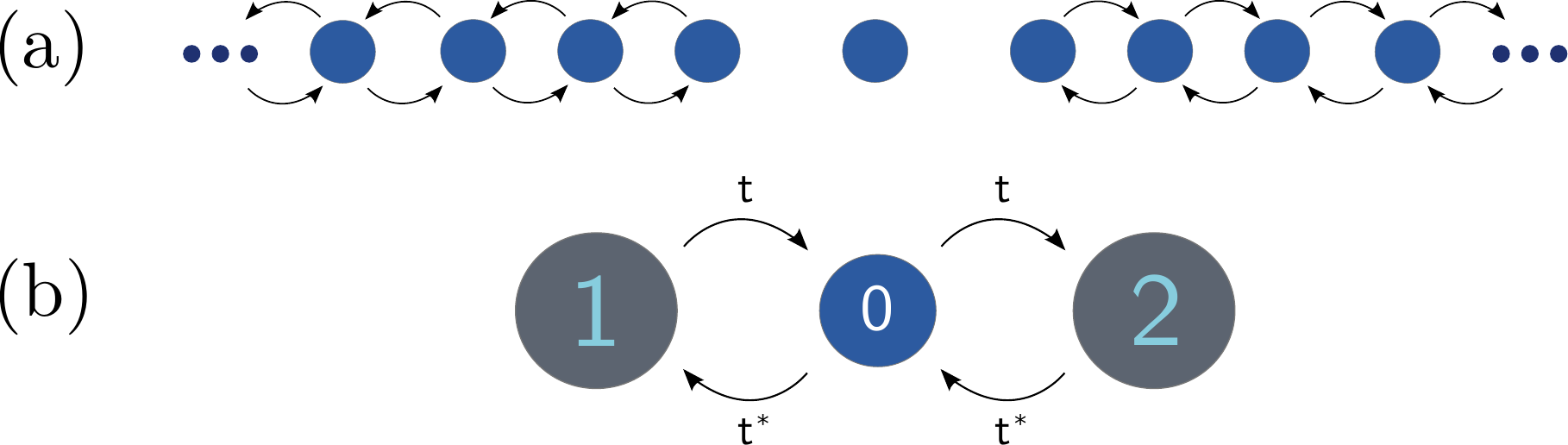}
 \caption{Infinite linear chain pictured as the coupling of two semi-infinite 
 chains with a single site.}
 \label{fig:inf3}
\end{figure}
\end{center}

In this case the equations of motion go not only forward but also 
backward. The dressed Green's function of the central site now reads
\begin{equation}
 G_{0}=\frac{g_0}{1-2 |t|^2 \,g_0 \tilde G_1 } \,, \label{eq:Ginf3}
\end{equation}
where $\tilde G_1$ is given by Eq.~\eqref{eq:semiG11w}. It can be shown that the 
expressions \eqref{eq:Ginf2} and \eqref{eq:Ginf3}  are identical, as long as 
$\tilde G_1$ obeys Eq.~\eqref{eq:G11baskara} (with $g_1=g_0$), which is indeed the 
case here. Replacing expression \eqref{glivre} for $g_0$ into 
Eq.~\eqref{eq:Ginf3} one obtains
\begin{equation}
 G_{0}=\frac{-i}{2|t|} \frac{1}{\sqrt{1-\left( \frac{\omega-  \epsilon_0 + 
i\eta}{2|t|} \right)^2}} \,. \label{eq:Ginf3omega}
\end{equation}

In Eq.~\eqref{eq:Ginf3omega} we can see that the resulting Green's function of the infinite chain has 
a square root singularity at $\omega - \varepsilon_0 = 2t$. The infinitesimal $\eta$ 
contributes to a softening around the singularity. For values $\omega - \varepsilon_0 < |2t|$,
the Green's function is in essence purely imaginary, with roughly the 
profile of an inverse of the semicircle we have seen in Fig.~\ref{fig:semiReIm}
however, with the presence of singularities at the band edges $\omega - \varepsilon_0 = |2t|$. These 
asymmetric spikes are a hallmark of low-dimensional systems (known as van 
Hove singularities), and indicate the presence of a flat dispersion curve with large 
accumulation of states. These singularities have effects on the structural, electrical and 
optical properties of solids and nanostructured materials, such as carbon nanotubes.
The density of states of the inner site, obtained with the imaginary 
part of the Green's functions Eq.~\eqref{eq:Ginf2} or Eq.~\eqref{eq:Ginf3}, 
is plotted in Fig.~\ref{gr:inf}. 

\begin{center}
\begin{figure}[ht]
 \centering\includegraphics[width=\columnwidth]{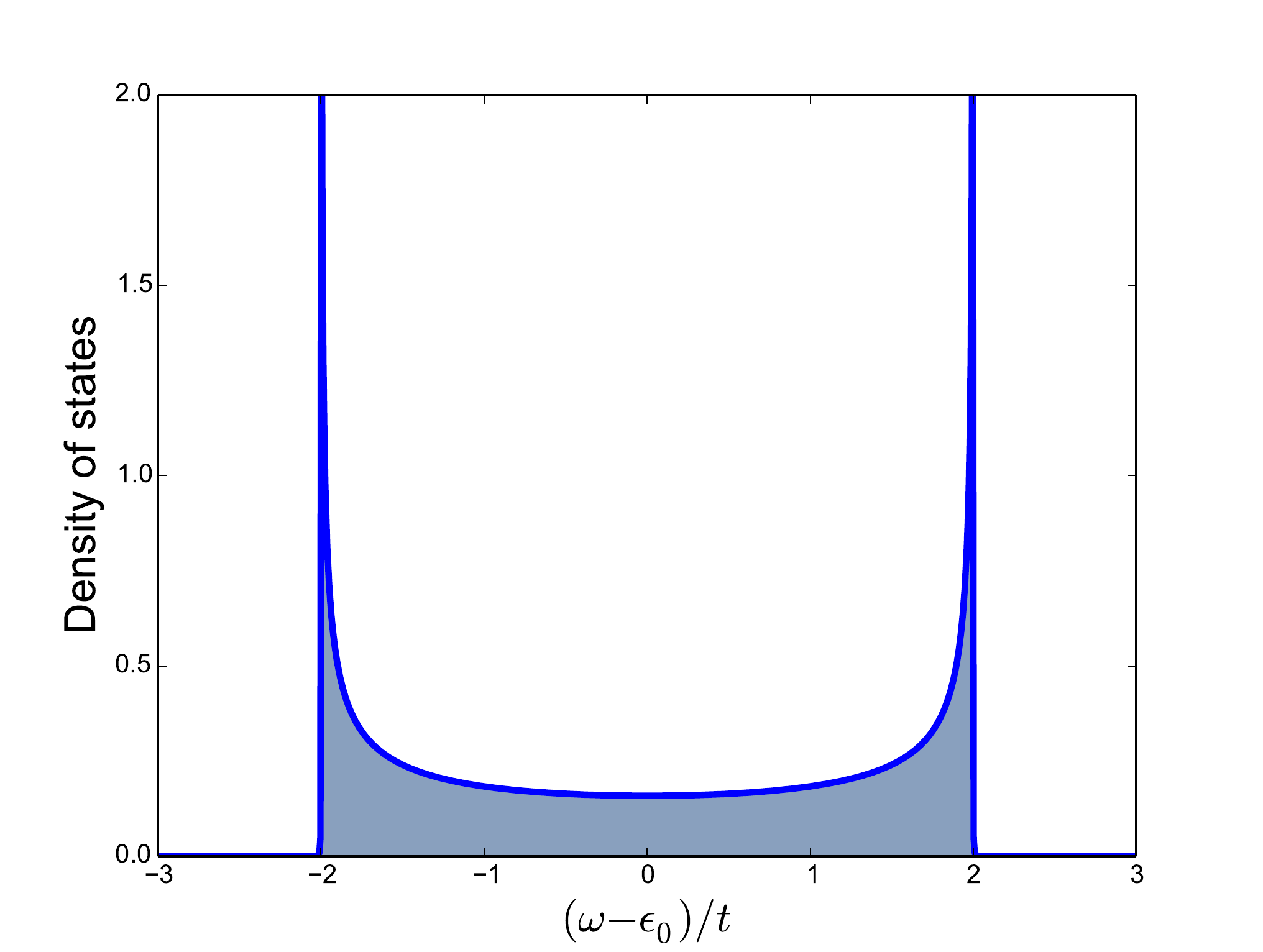}
 \caption{Density of states, Eq.~\eqref{eq:DOS}, of an infinite linear chain, obtained by merging two 
semi-infinite Green's functions. At the band edges we have a large accumulation 
of states, due to a flat band structure. These spikes are characteristic of  
low-dimensional periodic systems, and are known as van Hove singularities.}
 \label{gr:inf}
\end{figure}
\end{center}

In source code \ref{code:inf} (see Appendix), 
we have illustrated how to obtain the graph of Fig.~\ref{gr:inf} using the 
\texttt{Julia} programming language. For an introductory course in \texttt{Julia}, 
please see Ref.\cite{GersonJulia}.

\subsection{Three-site chain: a recipe for recursion}\label{sec:3sitios}

Let us now apply the equation-of-motion technique to a linear chain composed of 
three sites, shown in Fig.~\ref{fig:3sites}. Although it may appear as just 
another application of  Eq.~\eqref{EOMcin}, these equations will set our 
paradigm for the \textit{surface-bulk recursive Green's function} method 
presented in Sec.~\ref{sec:RGF3}. For the widely-used \textit{surface 
Green's function},  this 3-site model is revisited briefly, however 
special attention is required by the \textit{surface-bulk} method that will 
be presented.
\begin{center}
\begin{figure}[ht]
 \centering\includegraphics[height=2cm]{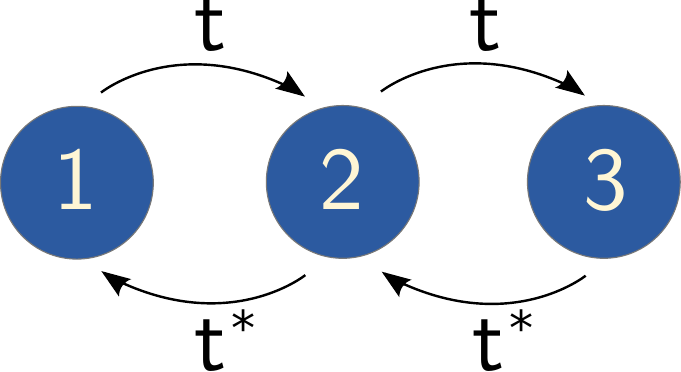}
 \caption{Three-site chain with nearest-neighbors hoppings $t$ and $t^*$. 
 The visualization of the sites may help writing of the equations of motion, 
 making it easier and mechanic. The equations of motion of this system will 
 play an important role for the recursive methods presented later on. } 
 \label{fig:3sites}
\end{figure}
\end{center}

Let us assume that our three-site chain is described by the non-interacting 
Hamiltonian
\begin{equation}
 H = \varepsilon_{0} \sum_{i=1}^{3} n_i + t (c^{\dagger}_2 
c_1 + c^{\dagger}_3 c_2) + t^*(c^{\dagger}_1 c_2 + c^{\dagger}_2 c_3)\,.
\end{equation}

From the local potential term of the Hamiltonian above, we see that  the 
undressed Green's functions \eqref{glivre} can be written as  
$g_i^r=(\omega - \varepsilon_0 + i\eta)^{-1}$. We now will write the EOM 
for the dressed Green's function $G^r_i$, and for the non-diagonal 
propagators $G^r_{ij}$ that connect the sites  $i$ and $j$. We will omit 
the energy dependence  ($\omega$) and the index $r$, for simplicity.

\paragraph{Green's function of site 1: $G_{11}$}---
Let us now calculate the Green's function of the first site of the three-site 
system, according to Eq.~\eqref{EOMcin}. Schematically we see in  
Fig.~\ref{fig:3sites} that the site  $1$ couples to site $2$ via a non-diagonal 
propagator  $G_{21}$ (where the subindex describes the propagator ``from the 
site $2$ to the site $1$''),
\begin{equation}
\boxed{ G_{11}=g_1 + g_1\, t \, G_{21} \label{eq:G11}   } \,.
\end{equation}

One way of visualizing how it works is first to identify the first neighbor of 
the site in question (see Fig.~\ref{fig:3sites}), the direction of the  
hopping, and the corresponding propagator $G_{kj}$, keeping in mind 
that the last index $j$ of the non-diagonal propagator has to be the same as 
the one of the Green's function  $G_{ij}$ under consideration.

The non-diagonal propagators that point to the first site are
\begin{align}
 G_{21}&=g_2 \, t^* \, G_{11} + g_2 \, t \, G_{31} \label{eq:3sitesG21}\\
 G_{31}&=g_3 \, t^* \, G_{21} \,. \label{eq:3sitesG31}
\end{align}

Inserting Eq.~\eqref{eq:3sitesG31} into \eqref{eq:3sitesG21}, we obtain 
\begin{align}
 G_{21}&=g_2 \, t^* \, G_{11} + g_2 \, t \, ( g_3 \, t^* \, G_{21} ), \\
 G_{21}&=\frac{g_2 \, t^* \, G_{11}}{(1 - g_2 \, t \, g_3 \, t^* )}.
 \label{eq:G21}
\end{align}

Using the result of Eq.~\eqref{eq:G21} in \eqref{eq:G11}, we can eliminate 
$G_{21}$ to obtain the dressed Green's function for the first site as:
\begin{equation}
 G_{11}=\frac{g_1}{ 1 - \displaystyle\frac{g_1\, t \, g_2 \, t^*}{1 
- g_2\,t\,g_3\,t^*}} \label{3G11} \,.
\end{equation}
 
\paragraph{Green's function of site 2: $G_{22}$}---
Applying the practical scheme discussed above we can write an expression for 
the central Green's  function as
\begin{equation}
\boxed{G_{22}=g_2 + g_2\, t^* \, G_{12} + g_3\, t \, G_{32}} \;. 
\label{eq:G22}
\end{equation}

Since there are only three sites, the expressions for propagators pointing 
to site 2 are
\begin{align}
 G_{12}&=g_1 \, t \, G_{22},  \\
 G_{32}&=g_3 \, t^* \, G_{22} \,.
\end{align}
These expressions are inserted into Eq.~\eqref{eq:G22} to obtain the 
local dressed Green's function of site $2$,
\begin{equation}
G_{22}=\frac{g_2}{ 1 - g_2\, t^* \, g_1 \, t - g_2\, t\, g_3\, t^*} \,. \label{3G22}
\end{equation}

\paragraph{Green's function of site 3: $G_{33}$}---
The equation of motion for the local dressed Green's function of site $3$ gives us 
\begin{equation}
    \boxed{G_{33}=g_3 + g_3\,t^*\,G_{23}} \,. \label{eq:G33}  
\end{equation}

To obtain a closed expression for $G_{33}$ we can either work on the EOM for 
the $G_{23}$ or just make the replacement $1\rightarrow 3$, $3\rightarrow 1$ 
and $t\rightarrow t^*$ in Eq.~\eqref{3G11}. The resulting expression is 
\begin{equation}
 G_{33}=\frac{g_3}{ 1 - \displaystyle\frac{g_3\, t^* \, g_2 \, t}{1 
- g_2\,t^*\,g_1\,t}} \label{3G33} \,.
\end{equation}

So far these examples not only provided us the opportunity to exercise the 
method but also introduced the boxed expressions \eqref{eq:G11}, \eqref{eq:G22} e 
\eqref{eq:G33}, fundamental to the technique developed in Sec.~\ref{sec:RGF3} for infinite chains.

\section{Recursive Green's function}

\subsection{Surface Green's functions decimation}\label{sec:RGFs}

In early 80's, the investigation of surface and bulk properties of metals,
transition metals and semiconductors motivated the development of effective 
Hamiltonians and iterative techniques to obtain the density of states \cite{Guinea}. 
The recursive Green's functions (RGF) used computationally efficient 
decimation techniques from the numerical renormalization group,
simulating materials via effective layers \cite{LopezSancho}. 

The success of recursive Green's functions was boosted by simulation of
transport in materials, in particular in two-terminal ballistic transport.
The retarded and advanced Green's functions of the central device in a 
junction contain information to the calculation of transport properties 
such as the stationary current and conductivity, or transmission matrix.
In essence, the idea of dividing the material in layers, modelling it in a chain, 
is the spirit of the recursive Green's function method. We will illustrate 
this procedure using a linear chain of single-site orbitals and two forms 
of decimation: the most widely-used, the surface technique, and an alternative
version that stores information from the central sites.

Let us consider a three-site chain, as shown in Fig.~\ref{fig:surface}(a). 
We will basically follow the references \cite{LopezSancho,Lewenkopf}  
except for the fact that in our notation, the first site is labelled as $1$ instead of $0$, 
therefore every index will be shifted by one with respect to the ones in \cite{LopezSancho,Lewenkopf}. 
Again, for the first site we have the equations of motion
\begin{align}
    G_{11}&=g_1 + g_1\,t\,G_{21} \label{eq:G00}\\
    G_{21}&=g_2\,t\,G_{31}+g_2\,t^*\,G_{11} \label{eq:G10surf}\,.
\end{align}

By replacing \eqref{eq:G10surf} in \eqref{eq:G00}, we \textit{eliminate} 
the non-diagonal propagator $G_{21}$:
\begin{equation}
    (1-g_1\,t\,g_2\,t^*)G_{11}=g_1+g_1\,t\,g_2\,t\,G_{31} 
\label{eq:G00surf} \,.
\end{equation}

As a general rule, the non-diagonal propagator $G_{n1}$ relates first neighbors:
\begin{align}
    G_{21}&=g_2\,t\,G_{31}+g_2\,t^*\,G_{11} \nonumber\\
    G_{31}&=g_3\,t\,G_{41}+g_3\,t^*\,G_{21} \nonumber\\
    G_{41}&=g_4\,t\,G_{51}+g_4\,t^*\,G_{31} \nonumber\\
    \vdots \nonumber\\
    G_{n1}&=g_n\,t\,G_{n+1,1}+g_n\,t^*\,G_{n-1,1} \label{eq:Gn0}\,.
\end{align}

Writing analogous expressions of \eqref{eq:Gn0} for $G_{n-1,1}$ and $G_{n+1,1}$, 
and replacing back into Eq.~\eqref{eq:Gn0}, we obtain 
a recursive expression that eliminates the non-diagonal first-neighbors 
propagators leaving only non-diagonal second-nearest neighbors functions:
\begin{equation}   
G_{n1}=\frac{g_n\,t\,g_{n+1}\,t\,G_{n+2,1}+g_n\,t^*\,g_{n-1}\,t^*\,G_{n-2,1
}}{1-g_n\,t\,g_{n+1}\,t^* - g_n\,t^*\,g_{n-1}\,t} \label{eq:Gn0eff}\,.
\end{equation}

Rewriting Eq.~\eqref{eq:Gn0eff} in terms of new variables
\begin{align}
    \alpha_1&=t\,g\,t \label{eq:alpha} \\
    \beta_1&=t^*\,g\,t^* \label{eq:beta} \\
    \tilde\varepsilon_1&=\varepsilon + t\,g\,t^* \\
    \varepsilon_1&=\tilde\varepsilon_1 + t^*\,g\,t \,,
\end{align}
where all undressed functions $g_i=g$ are given by \eqref{glivre}, 
we arrive at a shorter recursion relation
\begin{equation}
    (\omega-\varepsilon_1+i\eta)G_{n1}=\alpha_1\,G_{n+2,1} + 
\beta_1\,G_{n-2,1} \label{eq:G2nd}\,.
\end{equation}

Starting from $G_{11}$, Eq.~\eqref{eq:G2nd} generates a recursion relation 
involving only non-diagonal second-nearest neighbors functions of 
\textit{odd} sites. The first iteration is 
Eq.~\eqref{eq:G00surf}, involving sites $1$ and $3$. 
Next, the non-diagonal $G_{31}$ relates sites $1$ and $5$, and so on, as follows:
\begin{align}
    (\omega-\varepsilon_1+i\eta)&G_{11}=\alpha_1\,G_{31} + 1 \label{eq:G11odd} \\
    (\omega-\varepsilon_1+i\eta)&G_{31}=\alpha_1\,G_{51} + \beta_1\,G_{11} \\
    (\omega-\varepsilon_1+i\eta)&G_{51}=\alpha_1\,G_{71} + \beta_1\,G_{31} \\
    &\qquad \vdots \nonumber \\
    (\omega-\varepsilon_1+i\eta)&G_{2n+1,1}=\alpha_1\,G_{2(n+1)+1,1} + \beta_1\,G_{2(n-1)+1,1} \nonumber\\
    (\omega-\varepsilon_1+i\eta)&G_{2n+1,1}=\alpha_1\,G_{2n+3,1} + \beta_1\,G_{2n-1,1}
    \label{eq:recodd}   \,.
\end{align} 

These equations (except for the first one) are analogous to the 
first-neighbors recursion, Eq.~\eqref{eq:Gn0}, since their 
\textit{equations have the same structure}. However, the variables 
$\alpha_1$, $\beta_1$, etc, contain implicitly the nearest neighbors of the 
original chain, mapping now into a chain with \textit{twice} the lattice constant, 
since we connect second-nearest neighbors \cite{Guinea}.

\textit{Starting from} Eq.~\eqref{eq:recodd}, we can now repeat the 
arguments described above, from Eq.~\eqref{eq:alpha} to \eqref{eq:recodd},
$x$ times. At each repetition we will obtain a larger effective system with not twice, 
but $2^x$ the lattice constant. 
This process is known as \textit{decimation}, where one encapsulates the numerous 
sites into a three-point recursion relation using renormalized parameters.
This procedure ultimately provides information about the infinite lattice.
After $x$ iterations, Eq.~\eqref{eq:G11odd} to \eqref{eq:recodd} read

\begin{align}
(\omega - \varepsilon_x^S + i\eta)&G_{11}=\alpha_x G_{31} + 1 \nonumber\\
(\omega-\varepsilon_x+i\eta)&G_{2^x\cdot 1 + 1,1}=\alpha_1\,G_{2^x \cdot 2 + 1,1} + \beta_1\,G_{2^x \cdot 0 + 1,1} \nonumber \\
&\qquad \vdots \nonumber \\
(\omega - \varepsilon_x + i\eta)&G_{2^x \cdot n+1,1}= \alpha_x G_{2^x\cdot (n+1)+1,1} \nonumber\\ 
&+ \beta_x G_{2^x \cdot (n-1)+1,1}, 
\nonumber 
\end{align}
for $n\geq 1$. The renormalized hoppings are smaller than the original $t$, since they are 
multiplied by the undressed $g$, as in Eq.~\eqref{eq:alpha} and \eqref{eq:beta}. Those read
\begin{align}
    \alpha_x&=\alpha_{x-1}g_{x-1}\alpha_{x-1} \\
    \beta_x &= \beta_{x-1}g_{x-1}\beta_{x-1} \\
    \varepsilon_x^S &=\varepsilon_{x-1} + \alpha_{x-1}g_{x-1}\beta_{x-1} \\
    \varepsilon_x &= \varepsilon_x^S + \beta_{x-1}g_{x-1}\alpha_{x-1}\,,
\end{align}
where $g=(\omega - \varepsilon_{x-1} + i\eta)^{-1}$. 
After $x$ iterations, we have that site $1$ is coupled to a chain of 
$2^x$ sites where the effective hopping parameter is much smaller. 
The decimation will stop when $||\alpha_x||$ and $||\beta_x||$ are 
sufficiently small. At this point $\varepsilon_x \approx \varepsilon_{x-1}$, 
$\varepsilon^S_x \approx \varepsilon^S_{x-1}$, and 
\begin{equation}
(\omega - \varepsilon_x^S + i\eta)G_{11} \approx1 \,.
\end{equation}
Thus we have an approximation to the local Green function from the surface 
site $1$, at the edge of the chain:
\begin{equation}
 G_{11} \approx \frac{1}{(\omega - \varepsilon_x^S + i\eta)} \,.
\end{equation}

To have a picture of the decimation procedure, we illustrated the 
iterations steps in Fig.~\ref{fig:surface}. Note that it is the reverse 
of the encapsulating mechanism of the infinite lattice into a finite chain,
shown in Fig.~\ref{fig:inf} and \ref{fig:inf3}.
We start with the three-site chain, shown in Fig.~\ref{fig:surface}(a), 
and eliminate $G_{21}$, represented 
in the figure by the site $2$ in lighter color. 
In the first iteration, we add two interstitial sites, growing the lattice to 
$5$, shown in Fig.~\ref{fig:surface}(b). 
Next, we eliminate the \textit{even} non-diagonal functions, storing 
the information of the new sites into parameters $\alpha$, $\beta$, 
$\tilde\varepsilon$ and $\varepsilon$. With these renormalized parameters 
one can simulate a chain that grows exponentially fast 
keeping the three-point structure of Eq.~\eqref{eq:Gn0}. 

\begin{center}
\begin{figure}[ht]
 \centering\includegraphics[width=6cm]{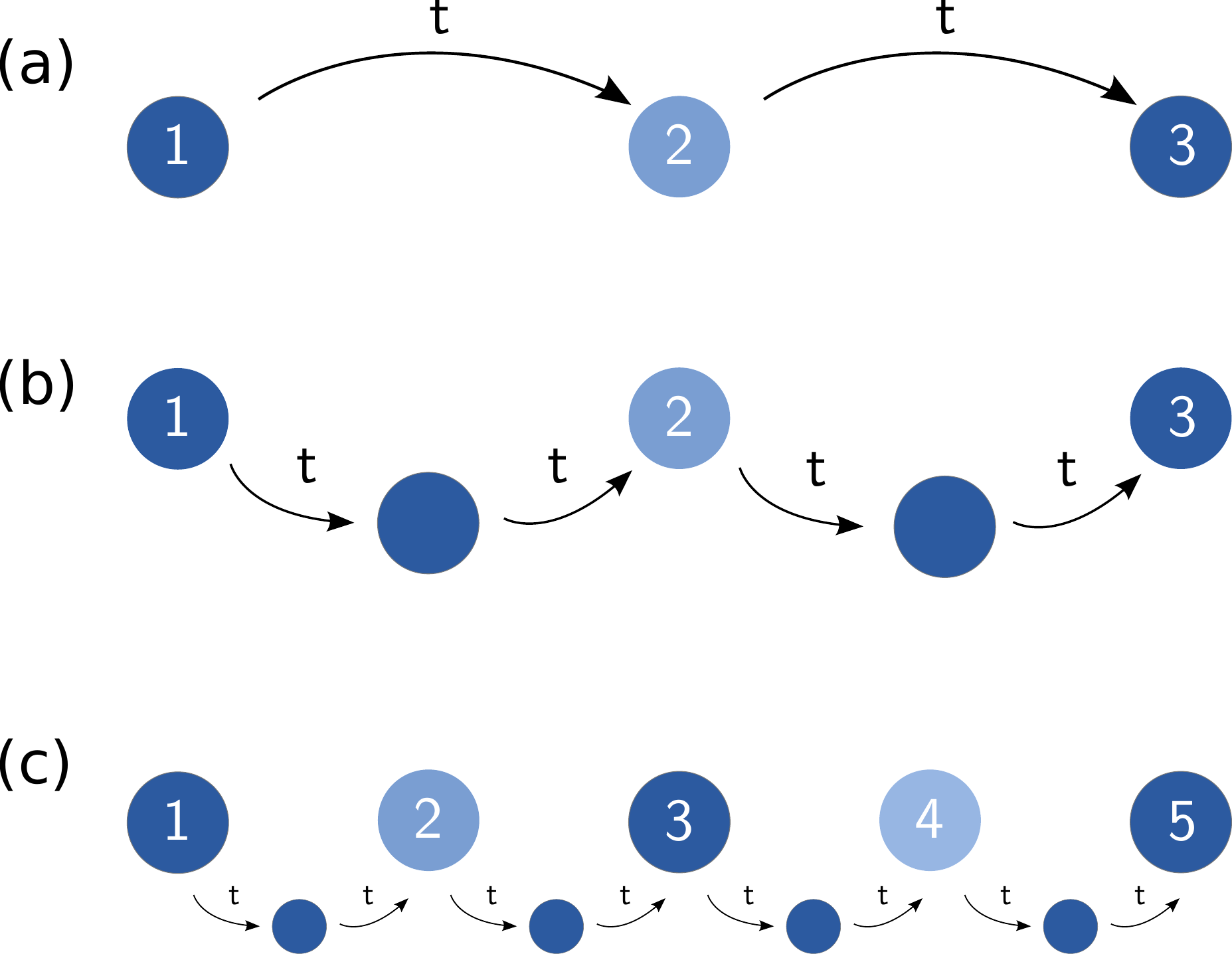}
 \caption{Possible interpretation of the decimation steps in the surface
 recursive Green's function. (a) 3-site chain, where the non-diagonal even 
 Green's  function from site $2$ is eliminated from the equations of motion,
 shown in a lighter color. 
 (b) Insertion of $2^1=2$ new sites, which will be included in a renormalization 
 of the hoppings.
 (c) In the next iteration, $2^2=4$ interstitial sites are inserted and the even
 non-diagonal propagators to the surface site, related to sites $2$ and $4$ 
 (in lighter color) will be eliminated. The idea is to keep the three-site chain 
 by renormalizing the hoppings and local energy of the first site. } 
 \label{fig:surface}
\end{figure}
\end{center}

The surface RGF is widely used in transport simulation with several 
applications \cite{Nardelli,Lewenkopf,Cuevas} with sophistications
\cite{Multiterminal}. In the next section we will present an alternative version,
capable to access the Green's functions of the edge and bulk at once, possibly
finding usefullness in topological insulators.\footnote{In fact, within 
the surface approach, it is possible to determine the bulk Green's 
function. One can consider an additional site and couple it from the 
left and from the right with semi-infinite chains, as we have shown in Fig.~\ref{fig:inf}
in Sec.~\ref{sec:infinite}. To this, one should 
first determine the surface GF from both sides, which usually are 
identical. However, they can differ for instance in topological systems,
where each side has its own chirality, or for asymmetric leads in transport devices.}

\subsection{Surface-bulk Recursive Green's function decimation} \label{sec:RGF3}

Another form of RGF, which we first present here, is based in the 3-site 
local GF, already introduced in Sec.~\ref{sec:3sitios}. The decimation is similar to 
the surface procedure, we will insert interstitial sites at each 
iteration. \textit{The difference is in which functions we eliminate in the 
hierarchy of equation of motions} and in the recursive model.

Although the equation of motion (EOM) procedure is quite mechanic, we will exemplify how the 
decimation develops in the first iteration of the surface-bulk RGF. By now the reader can 
probably jump into the effective equations, we elaborate them for the 
sake of 
clarity. 

Let us add two sites $a$ and $b$ to the 3-site chain, shown in Fig.~\ref{fig:5sites}:
\begin{center}
\begin{figure}[ht]
 \centering\includegraphics[height=2cm]{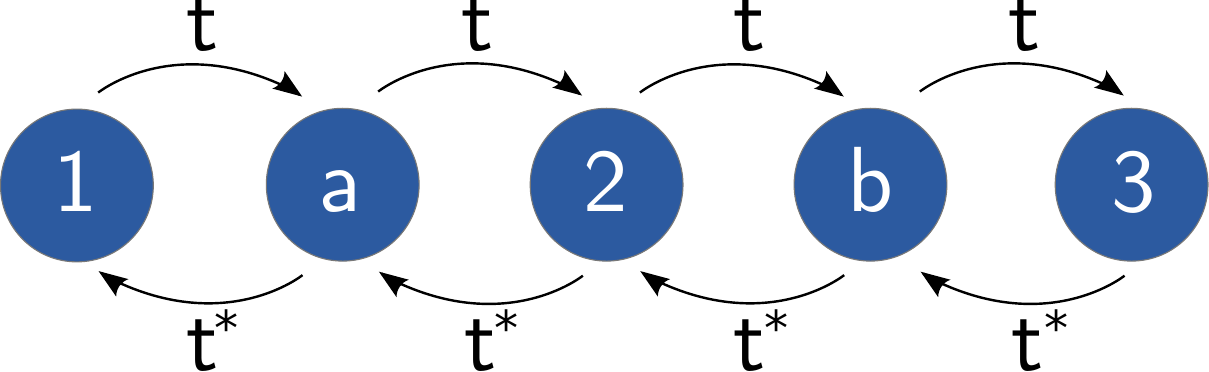}
 \caption{Illustration of the first decimation step, where we inserted interstitial sites $a$ 
 and $b$ in the three-site chain. } 
 \label{fig:5sites}
\end{figure}
\end{center}

For $5$ sites, the equations are more numerous and the surface solution will be 
more intrincate. We will examine three sites, the edges and the central site.

For the first site of Fig.~\ref{fig:5sites} we know that
\begin{align}
    G_{11}&=g_1 +g_1\,t\,G_{a1} \label{5:G11} \\
    G_{a1}&=g_a\,t^*\,G_{11} + g_a\,t\,G_{21} \label{Ga1} \,.
\end{align}

By replacing \eqref{Ga1} in \eqref{5:G11}, we eliminate the non-diagonal function $G_{a1}$
\begin{equation}  
G_{11}=\frac{g_1}{(1-g_1\,t\,g_a\,t^*)}+\frac{g_1\,t\,g_a\,t}{(1-g_1\,t\,
g_a\,t^*)}\,G_{21} \,. \label{eq:G11a121}
\end{equation}

Eq.~\eqref{eq:G11a121} can be rewritten in the form of the 
Eq.~\eqref{eq:G11} 
\begin{equation}
 G_{11}=\tilde{g_1} + \tilde{g_1}\,\tilde{t}\,G_{21}    \,,
\end{equation}
using the renormalized quantities 
\begin{equation}
    \tilde{g_1}=\frac{g_1}{(1-g_1\,t\,g_a\,t^*)} \quad \textrm{and} \quad 
    \tilde{t}=t\,g_a\,t \label{eq:g1tilde}  \,.
\end{equation}

Note that the edge propagator  $G_{11}$ corresponds to Eq.~\eqref{3G11},
\begin{equation}
 G_{11}=\frac{\tilde{g}_1}{\left[ 1 - \displaystyle\frac{\tilde{g}_1\, t 
\, \tilde{g}_2 \, t^*}{1 - \tilde{g}_2\,t\,\tilde{g}_3\,t^*} \right]} \,, \label{eq:G11RGFeff}
\end{equation}
with the undressed \textit{effective} functions $\tilde{g_2}$ e 
$\tilde{g_3}$, which we will derive, for completeness.

The Green's function for the central sites of Fig.~\ref{fig:5sites} has 
EOMs
\begin{align}
    G_{22}&=g_2 +g_2\,t^*\,G_{a2} + g_2\,t\,G_{b2} \label{5:G22}, \\
    G_{a2}&=g_a\,t\,G_{22} + g_a\,t^*\,G_{12} \label{Ga2}, \\
    G_{b2}&=g_b\,t\,G_{32} + g_b\,t^*\,G_{22} \label{Gb2} \,.
\end{align}

Eliminating the Green's functions \eqref{Ga2} and \eqref{Gb2},
we obtain Eq.~\eqref{eq:G22},
\begin{equation}
 G_{22}=\tilde{g_2} + \tilde{g_2}\,\tilde{t}^*\,G_{12} + 
\tilde{g_2}\,\tilde{t}\,G_{32}   \,,
\end{equation}
where we used the renormalized Green's function
\begin{equation}
    \tilde{g_2}=\frac{g_2}{(1-g_2\,t^*\,g_a\,t-g_2\,t\,g_b\,t^*)}  
\label{eq:g2tilde} \,.
\end{equation}

In Eq.~\eqref{eq:g2tilde}, $\tilde{t}^*=t^* g_a t^*$ e $\tilde{t}=t\,g_b\,t$,
considering undressed propagators $g_a=g_b$.

Finally, the Green's function for the last site of Fig.~\ref {fig:5sites} 
obeys the following equations,
\begin{align}
    G_{33}&=g_3 +g_3\,t^*\,G_{b3} \label{5:G33} \\
    G_{b3}&=g_b\,t\,G_{33} + g_b\,t^*\,G_{23} \label{Gb3} \,.
\end{align}

Comparing these expressions with \eqref{eq:G33}, we will 
consider $\tilde{t}^*=t^* g_b t^*$ in the renormalization of $g_3$
\begin{equation}
   \tilde{g_3}=\frac{g_3}{(1-g_3\,t^*\,g_b\,t)} \,. \label{eq:g3tilde} 
\end{equation}

In this five-site example we explicited the first step of the \textit{decimation 
recursion based on the three-site system}. This procedure is different 
from the surface Green's function approach, since we \textit{kept the 
three local propagators, eliminating the non-diagonal ones}. 
Figure~\ref{fig:35sites} illustrates the renormalization of the interactions and 
the mapping of the five-site chain onto the effective three-site one.
\begin{center}
\begin{figure}
 \centering\includegraphics[width=6cm]{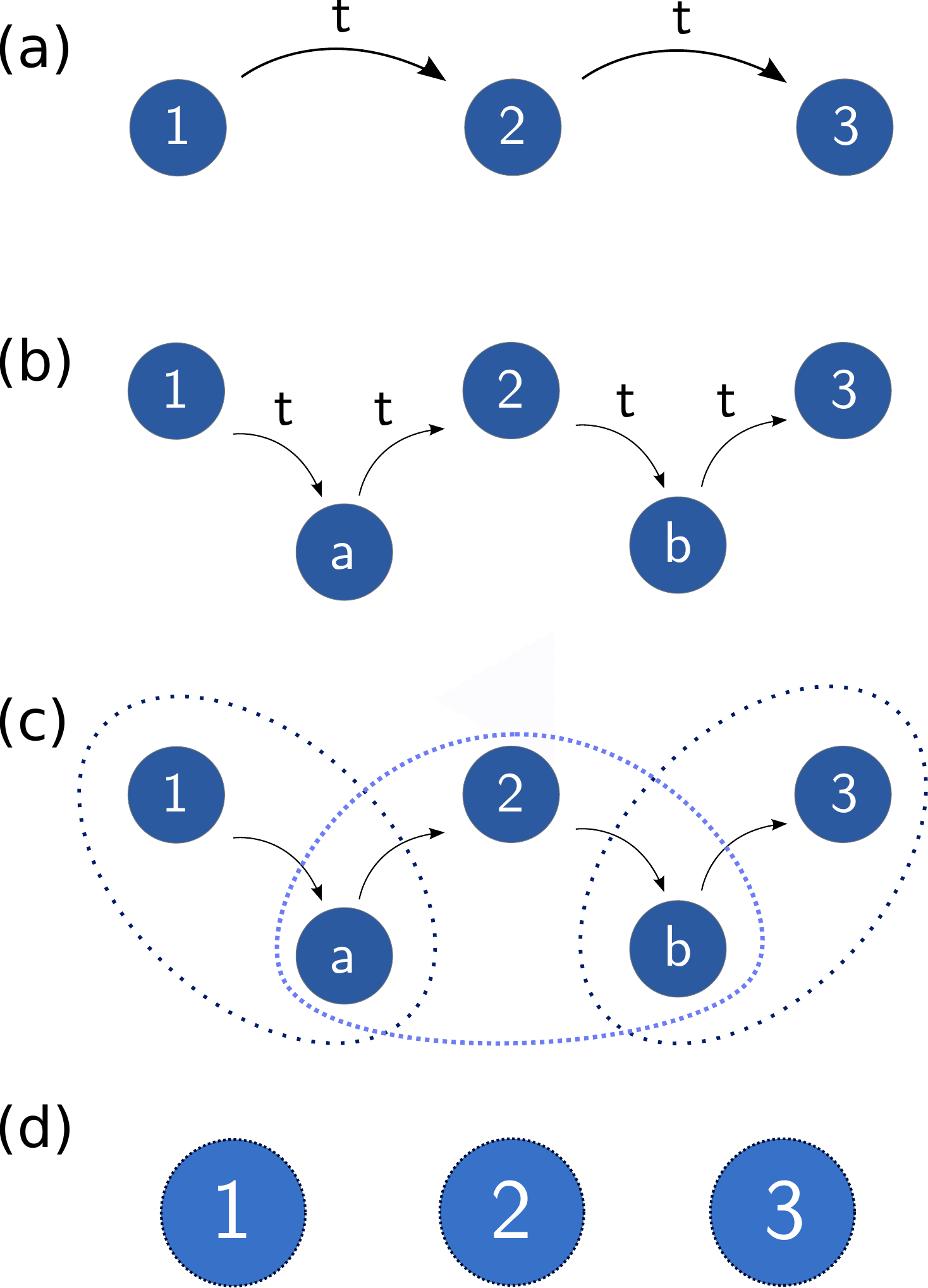}
 \caption{(a) e (b) Illustration of the iterative process of 
adding interstitial sites, representing the growth from a three-site 
to a five site chain. Panels (c) e (d) illustrate the recursive  
procedure of encapsulating the new sites to obtain the effective 
three-site system.} 
 \label{fig:35sites}
\end{figure}
\end{center}

In Fig.~\ref{gr:rgf5917}, we plot the imaginary part of the retarded Green's 
function, associated with the density of states, of the surface site 1, $\rho_{11}(\omega)$.
As the decimation procedure is carried, the number of peaks grows 
with the number of sites. The correspondent source code is presented in the
Appendix. 

\begin{center}
\begin{figure}[ht!]
 \centering\includegraphics[width=\columnwidth]{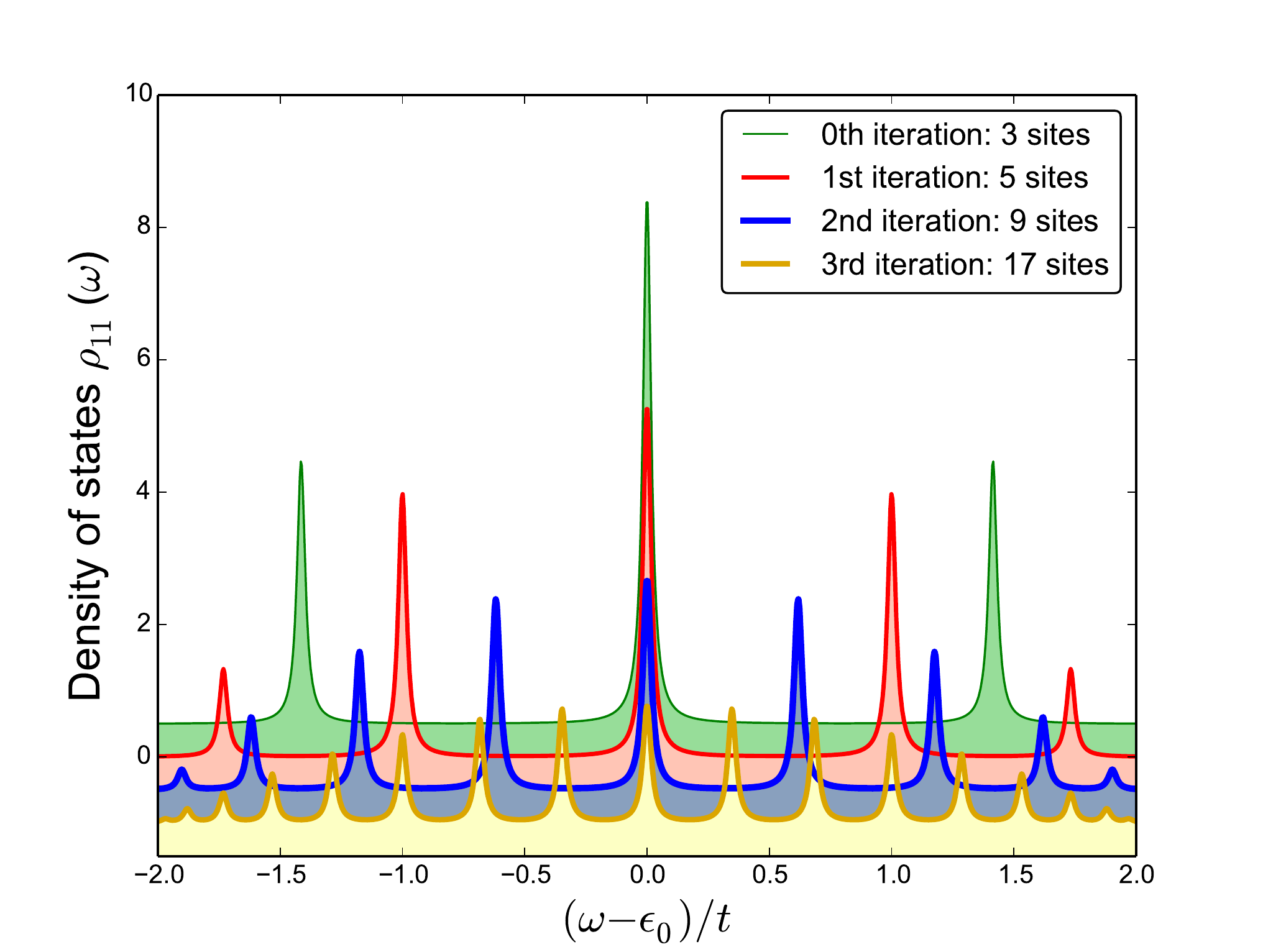}
 \caption{Density of states of the surface site at each step $x$ of the decimation, 
 showing the growth of the chain (as $2^{x}+1$) in the number of peaks. 
 We have shifted the curves vertically and set a large $\eta=0.02$ 
 (\textit{i.e.}, broadening of the peaks)  for better visualization. The 
algorithm is shown in the Appendix, source code \ref{RGF1D}, 
which simulates the semiinfinite chain. } 
 \label{gr:rgf5917}
\end{figure}
\end{center}


\subsubsection{Semi-infinite lattice} \label{sec:semiinf}

The \textit{surface-bulk RGF} decimation technique detailed in Sec.~\ref{sec:RGF3} is 
an alternative to the widespread \textit{surface} method that automatically 
delivers information about the central site. However, both methods scale 
exponentially with the number of iterations and are easily extended to 
two-dimensions via a matrix representation. Here we chose to ellaborate better 
how the proposed surface-bulk decimation works in practice.

\begin{center}
\begin{figure}[ht]
 \centering\includegraphics[width=\columnwidth]{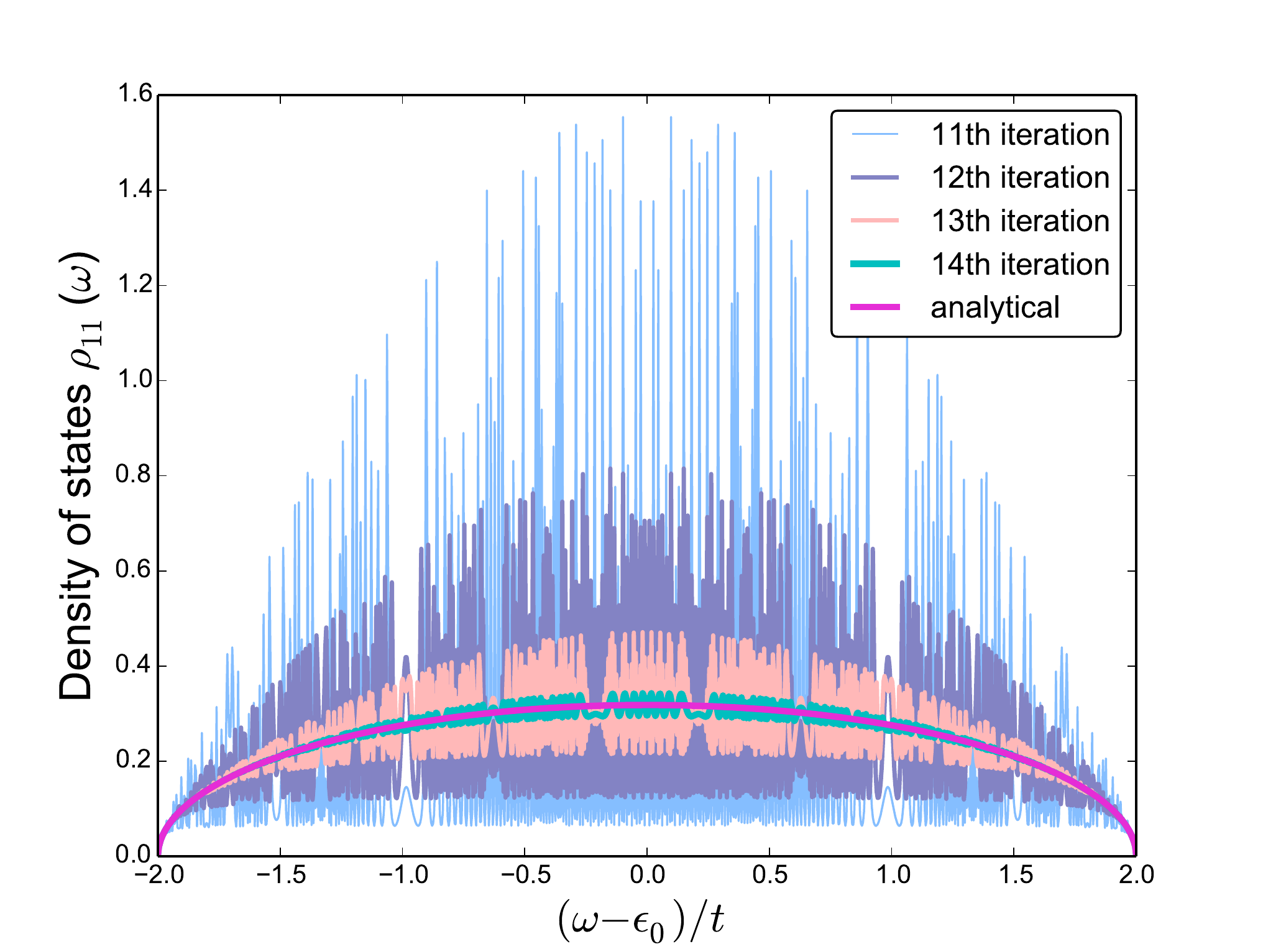}
 \caption{Density of states of the semi-infinite 
 linear chain evaluated with the RGF decimation procedure of Sec.~\ref{sec:RGF3} 
 and the analytical result of Eq.~\eqref{eq:dossemi}. In Fig.~\ref{gr:rgf5917} we showed 
 the first steps, here we plot from the 11th to 14th iteration, which exhibit several peaks. 
 For $\eta=10^{-4}$ the numerical RGF recovers the analytical expression around 16 steps, 
 $\approx66000$ sites.}
 \label{fig:semiinfingraf}
\end{figure}
\end{center}

We implemented the surface-bulk RGF algorithm in \texttt{Julia}. 
The source code \ref{RGF1D} (see Appendix) uses the recursive 
method to evaluate the surface density of states of a semi-infinite linear chain. 
The results of few steps are plotted in Fig.~\ref{gr:rgf5917} 
and Fig.~\ref{fig:semiinfingraf}.

\subsubsection{The ladder}

In order to approach two-dimensional materials, a generalization of the RGF 
decimation technique is usually performed by slicing a region (central device 
or lead) in layers, from which the surface algorithm follows \cite{Guinea}.
In two dimensions it is convenient to adopt a matrix representation of our Green's 
functions and hoppings.

We will approach this generalization in the simplest 2D example of a ladder, 
where we couple two 3-site chains vertically, as shown in Fig.~\ref{fig:ladder}. 
We will take as a convention a hopping $t$ to the right and upwards, 
and $t^*$ to the left or downwards. Each site will be indexed by its column (layer) $i$ 
and row $j$. We need to obtain the propagators $G_{ij,i'j'}$.
\begin{center}
\begin{figure}[ht]
 \centering\includegraphics[width=4cm]{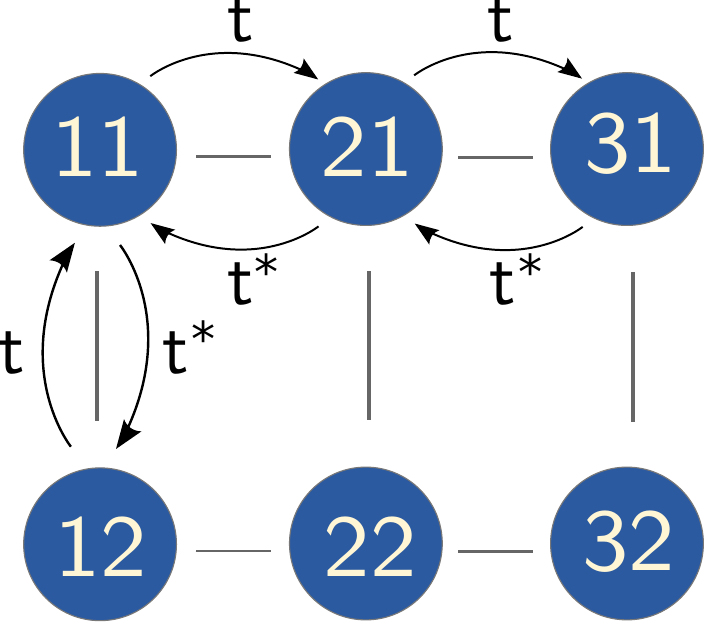}
 \caption{Generalization of the 3-site chain to a 2D design, which we refer as 
``ladder''. The new site indexes $ij$ correspond to the column $i$ and row 
$j$.} 
 \label{fig:ladder}
\end{figure}
\end{center}

Let us consider now displacements both on the horizontal as well as in the 
vertical direction. For example, the electron in the $11$ site can visit the two 
first neighbors $21$ or $12$ (see Fig.~\ref{fig:ladder}). The equation of motion of the 
$G_{11,11}$ site will exhibit then a self contribution $11$ and two non-diagonal 
propagators $G_{21,11}$ e $G_{12,11}$. The EOMs of this \textit{first column 
$i=1$} are
 \begin{align}
  G_{11,11} &= g_{11} + g_{11}\,t\,G_{21,11} + g_{11}\,t^*\,G_{12,11} \\
  G_{12,12} &= g_{12} + g_{12}\,t\,G_{11,12} + g_{12}\,t\,G_{22,12} \\
     G_{11,12}&=g_{11}\,t^*\,G_{12,12} + g_{11}\,t\,G_{21,12}\\
     G_{12,11}&=g_{12}\,t\,G_{11,11} + g_{12}\,t\,G_{22,11} \,.
 \end{align}

Arranging these equations in matrix form, we obtain
\begin{widetext}
\begin{eqnarray}
\begin{pmatrix}
G_{11,11} & G_{11,12} \\
G_{12,11} & G_{12,12} 
\end{pmatrix}
&\!=\!&
\begin{pmatrix}
g_{11} & 0 \\
0 & g_{12} 
\end{pmatrix} +
\begin{pmatrix}
g_{11} & 0 \\
0 & g_{12} 
\end{pmatrix}
\begin{pmatrix}
0 & t^* \\
t & 0
\end{pmatrix}
\begin{pmatrix}
G_{11,11} & G_{11,12} \\
G_{12,11} & G_{12,12} 
\end{pmatrix} +
\begin{pmatrix}
g_{11} & 0 \\
0 & g_{12} 
\end{pmatrix}
\begin{pmatrix}
t & 0 \\
0 & t
\end{pmatrix}
\begin{pmatrix}
G_{21,11} & G_{21,12} \\
G_{22,11} & G_{22,12} 
\end{pmatrix}.\nonumber\\
\label{G1111}
\end{eqnarray}
\end{widetext}

Notice that Eq.~\eqref{G1111} corresponds only to the first slice (column $1$). 
Casting the left-hand side (l.h.s.) as $\mathbf{G}_{1}$ and the undressed function as $\mathbf{g}_{1}$, 
we can identify two hopping matrices, one from same-column sites $\mathbf{V}$, 
and one between columns $\mathbf{W}$:
\begin{equation}
    \mathbf{G}_{1} = \mathbf{g}_{1} + \mathbf{g}_{1}\cdot \mathbf{V} \cdot 
\mathbf{G}_{1} + \mathbf{g}_{1} \cdot \mathbf{W} \cdot \mathbf{G}_{21} \,. \label{eq:G1matrix}
\end{equation}

By isolating $\mathbf{G_1}$ we can write
\begin{equation}
    \mathbf{G}_{1} = \mathbf{\bar g}_{1} + \mathbf{\bar g}_{1} \cdot \mathbf{W} 
\cdot \mathbf{G}_{21} \,,
\label{eq:G1matrix1}
\end{equation}
where we have defined
\begin{equation}
 \mathbf{\bar g}_{1}=\left( \mathbf{I} - \mathbf{g}_{1}\cdot \mathbf{V} 
\right)^{-1}\mathbf{g}_{1},  \, \label{eq:G1Vmatrix}
\end{equation}
that represents the Gren's function of a single slice.

From  Eq.~\eqref{eq:G1matrix1}, we can identify that the same 3-site structure 
of Eq.~\eqref{eq:G11} is now recovered in matrix form. This is very 
convenient, since we will be able to implement decimation in two dimensions.

For the second slice (column $i=2$), we have
 \begin{eqnarray*}
  G_{21,21} &=& g_{21} + g_{21}\,t^*\,G_{11,21} + g_{21}\,t^*\,G_{22,21} \\
  && + g_{21}\,t\,G_{31,21} \\
  G_{22,22} &=& g_{22} + g_{22}\,t^*\,G_{12,22} + g_{22}\,t\,G_{21,12} \\
  && + g_{22}\,t\,G_{32,22}\\
  G_{21,22}&=& g_{21}\,t^*\,G_{11,22} + g_{21}\,t^*\,G_{22,22} + 
g_{21}\,t\,G_{31,22}\\
  G_{22,21}&=&g_{22}\,t^*\,G_{12,21} + g_{22}\,t\,G_{21,21} + 
g_{22}\,t\,G_{32,21}\,,
 \end{eqnarray*}
which is represented as
\begin{widetext}
\begin{eqnarray}
\begin{pmatrix}
G_{21,21} & G_{21,22} \\
G_{22,21} & G_{22,22} 
\end{pmatrix}
\!&=&\!
\begin{pmatrix}
g_{21} & 0 \\
0 & g_{22} 
\end{pmatrix}
+
\begin{pmatrix}
g_{21} & 0 \\
0 & g_{22} 
\end{pmatrix}
\begin{pmatrix}
0 & t^* \\
t & 0
\end{pmatrix}
\begin{pmatrix}
G_{21,21} & G_{21,22} \\
G_{22,21} & G_{22,22} 
\end{pmatrix} 
+
\begin{pmatrix}
g_{21} & 0 \\
0 & g_{22} 
\end{pmatrix}
\begin{pmatrix}
t^* & 0 \\
0 & t^*
\end{pmatrix}
\begin{pmatrix}
G_{11,21} & G_{11,22} \\
G_{12,21} & G_{12,22} 
\end{pmatrix}  \nonumber \\
&&+
\begin{pmatrix}
g_{21} & 0 \\
0 & g_{22} 
\end{pmatrix}
\begin{pmatrix}
t & 0 \\
0 & t
\end{pmatrix}
\begin{pmatrix}
G_{31,21} & G_{31,22} \\
G_{32,21} & G_{32,22} 
\end{pmatrix}.
\label{G2222}
\end{eqnarray}
\end{widetext}

Therefore we can also rewrite Eq.~\eqref{G2222} in the same form of Eq.~\eqref{eq:G22}, 
from the three-site formulas:
\begin{equation}
    \mathbf{G}_{2} = \mathbf{g}_{2} + \mathbf{g}_{2}\cdot \mathbf{V} \cdot 
\mathbf{G}_{2} + \mathbf{g}_{2} \cdot \mathbf{W^*} \cdot \mathbf{G}_{12} + 
\mathbf{g}_{2} \cdot \mathbf{W} \cdot \mathbf{G}_{32} \,.
\end{equation}

From the two identifications above we can perform a mapping to three effective sites, 
corresponding to these slices, shown in Fig.~\ref{fig:laddermap}. The 
decimation method applies, allowing the simulation \textit{e.g.}, of a stripe. 
\begin{center}
\begin{figure}[ht]
 \centering\includegraphics[width=4cm]{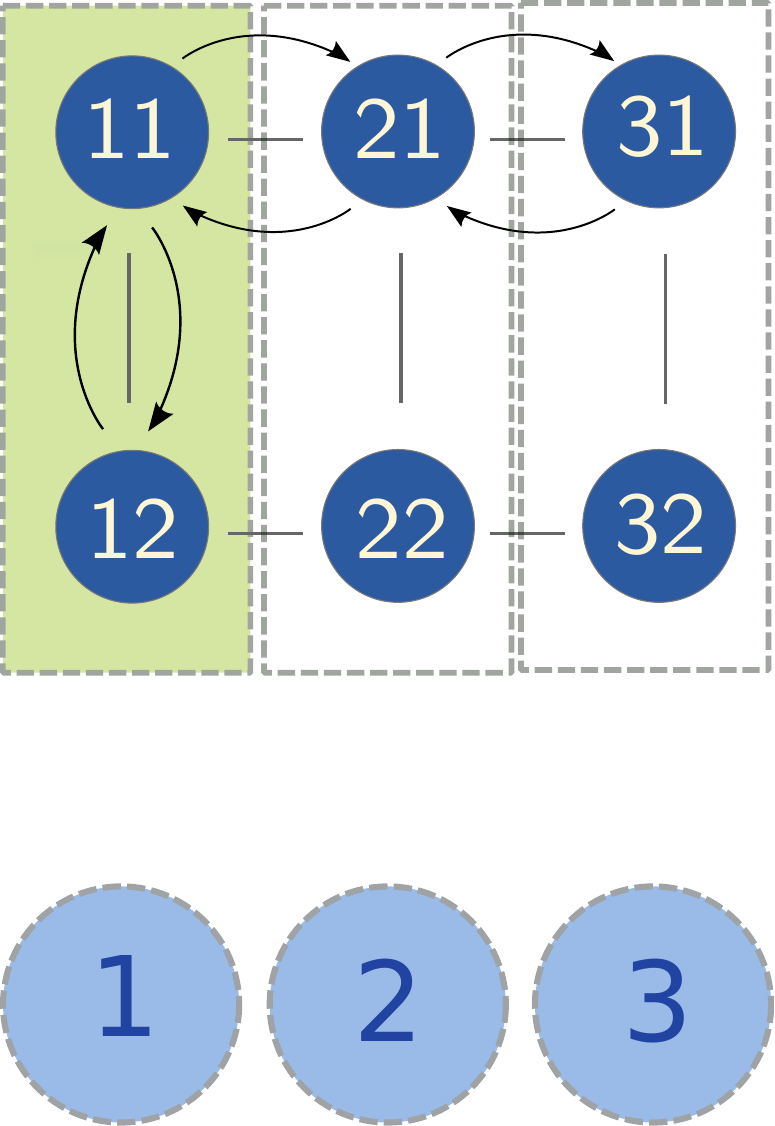}
 \caption{Mapping of the slices in 3 new effective sites.} 
 \label{fig:laddermap}
\end{figure}
\end{center}

The program in \texttt{Julia} to generate the results of the ladder is shown in the 
Appendix.

\begin{center}
\begin{figure}[ht]
 \centering\includegraphics[width=\columnwidth]{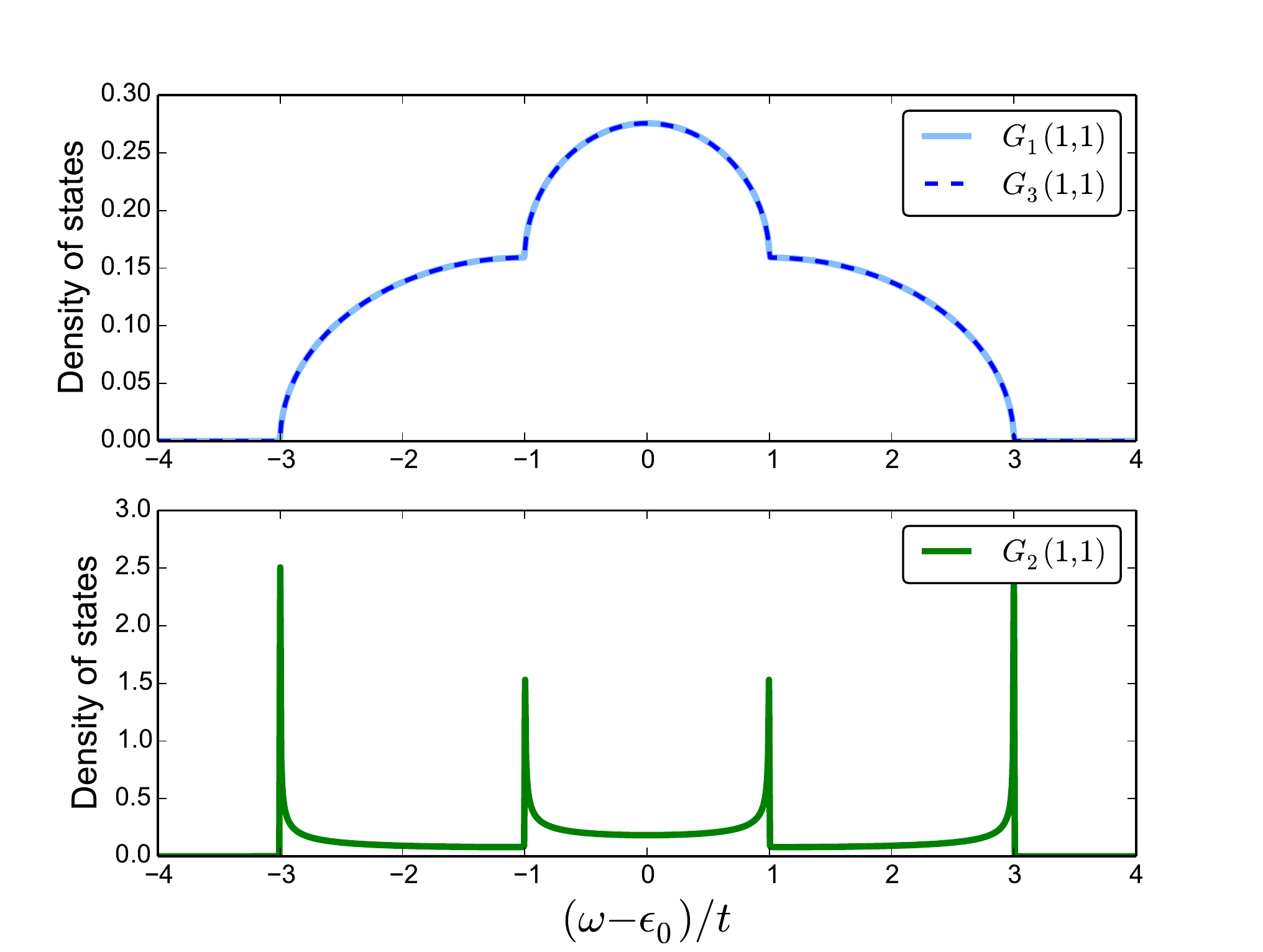}
 \caption{Density of states of the ``ladder'', an infinite stripe of width $L=2$.}
 \label{gr:dosladder}
\end{figure}
\end{center}

To go beyond the ladder, we can generalize $\mathbf{V}$ and $\mathbf{W}$ 
to bigger slices. These matrices will be larger but have a simple form, let us 
develop them. 

First note that, in a given slice, the electron can hop up or down a row. By our 
definitions (see Fig.~\ref{fig:ladder}), the down hopping is $t^*$, i.e., the 
hopping between $(i,j)$ e $(i,j+1)$, such as $11$ and $12$. Ordering the basis 
according to the row $j$, for the first column $i=1$ we have $\{11, 12, 13, 
\cdots \}$ (first index is $i=1$ and the second is $j=1,2,3, , \cdots$). The 
possible hoppings $\mathbf{V}$ in the first slice lead to a tridiagonal matrix 
with null diagonal, reflecting the fact that the hopping $\mathbf{V}$ takes the 
electron of the slice to different rows, the upper $(i,j+1)$ or lower one $(i,j-1)$ one:
\begin{equation}
\mathbf{V} = 
\begin{bmatrix}
0 & t^* & 0 & 0  & \cdots\\
t & 0 & t^* & 0  &\cdots\\
0 & t & 0 & t^*  &\cdots\\
0 & 0 & t & 0 &\cdots\\
\vdots & \vdots  & \vdots & \vdots & \ddots 
\end{bmatrix} \,.
\label{eq:V}
\end{equation}

For the $\mathbf{W}$ matrix, the hopping takes place between sites of 
different columns. Presently we deal with three effective sites, but as the 
decimation proceeds, the lattice will grow horizontally, forming a stripe.
In this process, notice that independently of the column $i$, automatically
\textit{all rows $j$ of the slice will be connected} since the slices will touch each other. 
For a given column $i=1$, for instance, with base order $\{11, 12, 13, \cdots \}$, 
where the second index is the row $j=1,2,3, \cdots$, every row is self-connected,
meaning that we have a diagonal matrix:
\begin{equation}
\mathbf{W} = 
\begin{bmatrix}
t & 0 & 0 & 0 & \cdots\\
0 & t & 0 & 0 & \cdots\\
0 & 0 & t & 0 & \cdots\\
0 & 0 & 0 & t & \cdots\\
\vdots & \vdots & \vdots & \vdots & \ddots 
\end{bmatrix} \,.
\label{eq:W}
\end{equation}

Therefore one can generalize the algorithm of the ladder to a stripe geometry, 
using the matrices \eqref{eq:V} and \eqref{eq:W} \footnote{To generalize the 
source code \ref{RGF2D} (Appendix) to a stripe, one should define a 
variable for the stripe size \texttt{Ly}, which in the case of the ladder is $\texttt{Ly=2}$.
The matrices \texttt{V} and \texttt{W} should be defined according to this size, 
\texttt{V = diagm(tv*ones(Ly-1),-1)+diagm(zeros(Ly))}
\texttt{+diagm(tv*ones(Ly-1),1)} and 
\texttt{W = tw*eye(Ly)}, where the command \texttt{eye} in \texttt{Julia} defines 
an identity matrix and \texttt{diagm} a diagonal matrix.}. In Fig.~\ref{gr:dosstripe} 
we plot the density of states of the bulk Green's function $G_2$ at the middle of 
the stripe, for different widths $L=2$ (ladder), $L=6$, and $L=128$. 

As we increase the width of the stripe, the behavior tends to the limit of an 
infinite square lattice, given by an analytic expression in terms an ellyptical 
function of the first kind \cite{Economou}. It exhibits a cusp at $\omega=0$, a 
logarithmic singularity characteristic of two-dimensional lattices. It is associated 
with critical saddle points in the two-dimensional band structure \cite{Callaway}.

\begin{center}
\begin{figure}[ht]
 \centering\includegraphics[width=\columnwidth]{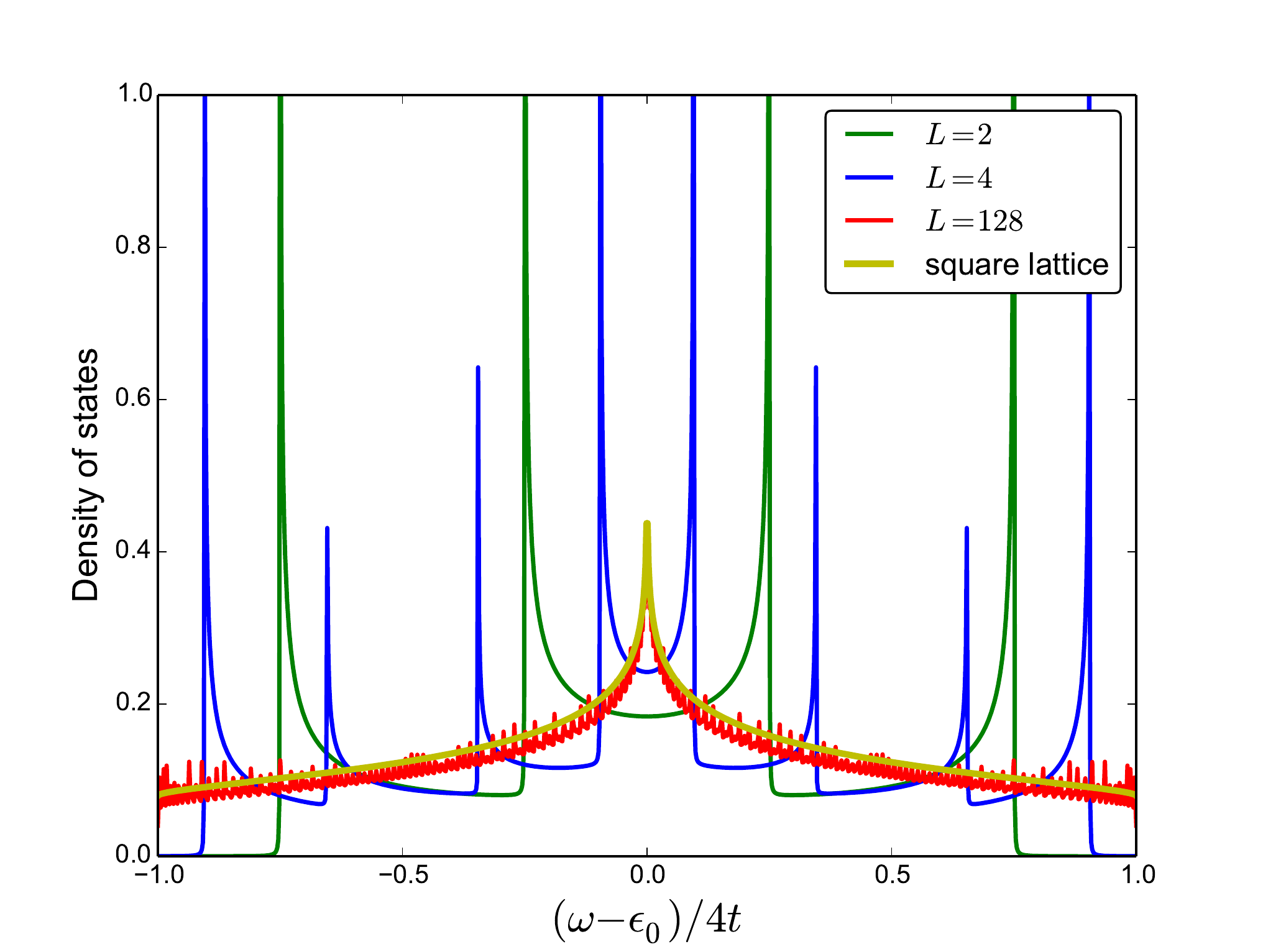}
 \caption{Local density of states of the bulk Green's function $G_2$ evaluated 
 inside a stripe of width $L$. We plotted the matrix element $(L/2,L/2)$, using $\eta=10^{-3}$. 
 The analytical result of the infinite square lattice \cite{Economou} is shown as a reference of the 
 asymptotic limit.}
 \label{gr:dosstripe}
\end{figure}
\end{center}

This last example illustrates the power of this technique in simulating 
finite lattices, which can go beyond the present regular chains to real  
nano or mesoscopic systems, such as electrodes, cavities, quantum dots and molecular junctions.

\section{Conclusions}

To conclude, we have presented a pedagogical introduction que the Green's 
function in the many-body formalism. Starting  with a general view of Green's 
functions, from the classical mathematical origin, going through the many-body 
definitions, we finally reached a practical application within the recursive 
Green's functions technique. For a young researcher, it is not easy to  grasp 
the whole power and at the same time, the tiny details of the numerical methods 
available.  Therefore we prepared this introduction based on 
simple condensed-matter  models with additional implementations in 
\texttt{Julia}, an open-source high-level language for scientific computing.

The \textit{surface-bulk} recursive Green's function is, to the best of our 
knowledge, a new proposal to the field, which brings an advantage in the 
investigation of topological materials, where one is interested in the edge 
and the bulk properties. Like the \textit{surface} approach, our 
\textit{surface-bulk} recursive Green's  function can be generalized to  other systems and
geometries \cite{Multiterminal,Lewenkopf}. We believe this material will be 
also useful for researchers unfamiliar with the Green's function method,  
interested in the new challenges of nanosciences and their implementations.


\acknowledgments
This work was partially supported by the Brazilian agencies CAPES, CNPq  and 
FAPEMIG. We would like to acknowledge Ginetom S. Diniz, Gerson J. Ferreira, and
Marcel Novaes for suggestions and careful reading.
\bibliographystyle{apsrev4-1}
\bibliography{ref.bib}

\appendix\section{Source codes}

\texttt{Julia} is a high-level, high-performance, easy-to-learn scientific 
language \cite{Bezanson}.
It is also an open-source project, licensed by MIT.
For an introductory course, please see for instance Reference \cite{GersonJulia}.

In source code \ref{code:inf}, we define a linearly spaced vector of energies 
using the command \texttt{linspace} and evaluate the undressed Green's function
from this vector. This shortened notation avoids additional and traditional use of 
the \texttt{for} loop for energies, which is inefficient, since the vector 
can be stored in memory at once, on the fly. If the amount of data to be stored 
is under the memory resources, vectorization of loops is a general recommended 
programming practice, since matrix and vector operations can be performed efficiently in \texttt{Julia}.
When we start evaluating more complex Green's functions, stored as large 
matrices, we return to the conventional loop of energies.

\begin{algorithm}[H]
\floatname{algorithm}{Source code}
\caption{Infinite chain}\label{code:inf}
\begin{algorithmic}[1]

\begin{Verbatim}[commandchars=\\\{\},frame=none,framesep=1.5ex,framerule=0.8pt,fontsize=\smaller]
\PY{c}{\PYZsh{} Julia programming language version 0.4.2}
\PY{c}{\PYZsh{} http://julialang.org/}

\PY{k}{using} \PY{n}{PyPlot}            \PY{c}{\PYZsh{} Matplotlib library}

\PY{n}{e0} \PY{o}{=} \PY{l+m+mf}{0.0}                \PY{c}{\PYZsh{} local site energy}
\PY{n}{eta} \PY{o}{=} \PY{l+m+mf}{1e\PYZhy{}4}              \PY{c}{\PYZsh{} positive infinitesimal}
\PY{n}{wmin} \PY{o}{=} \PY{o}{\PYZhy{}}\PY{l+m+mf}{2.0}\PY{p}{;} \PY{n}{wmax} \PY{o}{=} \PY{l+m+mf}{2.0} \PY{c}{\PYZsh{} energy range}
\PY{n}{Nw} \PY{o}{=} \PY{l+m+mi}{1000}               \PY{c}{\PYZsh{} number of energy points}
\PY{n}{w} \PY{o}{=} \PY{n}{linspace}\PY{p}{(}\PY{n}{wmin}\PY{p}{,}\PY{n}{wmax}\PY{p}{,}\PY{n}{Nw}\PY{p}{)} \PY{c}{\PYZsh{} vector of energies}
\PY{n}{g} \PY{o}{=} \PY{l+m+mf}{1.}\PY{o}{/}\PY{p}{(}\PY{n}{w}\PY{o}{\PYZhy{}}\PY{n}{e0}\PY{o}{+}\PY{n}{eta}\PY{o}{*}\PY{n+nb}{im}\PY{p}{)}    \PY{c}{\PYZsh{} undressed propagator }
\PY{n}{t} \PY{o}{=} \PY{l+m+mf}{1.0}                 \PY{c}{\PYZsh{} symmetric real hopping}

\PY{c}{\PYZsh{} Semi\PYZhy{}infinite chain analytic expression G\PYZus{}11}

\PY{n}{Gsemi} \PY{o}{=} \PY{p}{(}\PY{l+m+mf}{1.}\PY{o}{/}\PY{p}{(}\PY{n}{g}\PY{o}{*}\PY{l+m+mi}{2}\PY{o}{*}\PY{n}{t}\PY{o}{\PYZca{}}\PY{l+m+mi}{2}\PY{p}{)}\PY{p}{)}\PY{o}{.}\PY{o}{*}\PY{p}{(}\PY{l+m+mi}{1} \PY{o}{\PYZhy{}} \PY{n}{sqrt}\PY{p}{(}\PY{l+m+mi}{1}\PY{o}{\PYZhy{}}\PY{l+m+mi}{4}\PY{o}{*}\PY{n}{t}\PY{o}{\PYZca{}}\PY{l+m+mi}{2}\PY{o}{*}\PY{n}{g}\PY{o}{.}\PY{o}{\PYZca{}}\PY{l+m+mi}{2}\PY{p}{)}\PY{p}{)}

\PY{c}{\PYZsh{} Infinite chain analytic expression obtained}
\PY{c}{\PYZsh{} by joining two semi\PYZhy{}infinite chains}

\PY{n}{Ginf} \PY{o}{=} \PY{n}{Gsemi}\PY{o}{.}\PY{o}{/}\PY{p}{(}\PY{l+m+mi}{1}\PY{o}{\PYZhy{}}\PY{n}{Gsemi}\PY{o}{.}\PY{o}{\PYZca{}}\PY{l+m+mi}{2}\PY{o}{*}\PY{n}{t}\PY{o}{\PYZca{}}\PY{l+m+mi}{2}\PY{p}{)}

\PY{n}{xlabel}\PY{p}{(}\PY{l+s}{L\PYZdq{}}\PY{l+s}{Energy }\PY{l+s+si}{\PYZdl{}}\PY{l+s}{\PYZbs{}}\PY{l+s}{omega }\PY{l+s+si}{\PYZdl{}}\PY{l+s}{\PYZdq{}}\PY{p}{,} \PY{n}{fontsize}\PY{o}{=}\PY{l+m+mi}{20}\PY{p}{)}
\PY{n}{ylabel}\PY{p}{(}\PY{l+s}{\PYZdq{}}\PY{l+s}{Density of states}\PY{l+s}{\PYZdq{}}\PY{p}{,} \PY{n}{fontsize}\PY{o}{=}\PY{l+m+mi}{20}\PY{p}{)}
\PY{n}{axis}\PY{p}{(}\PY{p}{[}\PY{o}{\PYZhy{}}\PY{l+m+mi}{2}\PY{p}{,}\PY{l+m+mi}{2}\PY{p}{,}\PY{l+m+mi}{0}\PY{p}{,}\PY{l+m+mf}{1.4}\PY{p}{]}\PY{p}{)}
\PY{n}{plot}\PY{p}{(}\PY{n}{w}\PY{p}{,} \PY{p}{(}\PY{o}{\PYZhy{}}\PY{l+m+mf}{1.0}\PY{o}{/}\PY{n+nb}{pi}\PY{p}{)}\PY{o}{*}\PY{n}{imag}\PY{p}{(}\PY{n}{Ginf}\PY{p}{)}\PY{p}{,} \PY{n}{linewidth}\PY{o}{=}\PY{l+m+mf}{3.0}\PY{p}{)}
\end{Verbatim}

\end{algorithmic}
\end{algorithm}

Code \ref{RGF1D} uses the recursive method to evaluate the 
surface density of states of a semi-infinite linear chain. We use again 
the vectorized loop of energies \texttt{w} in the \texttt{linspace} command.
The explicit \texttt{for} loop runs the recursive decimation procedure 
for \texttt{16} steps. Equations \eqref{eq:g1tilde}, 
\eqref{eq:g2tilde} and \eqref{eq:g3tilde} are implemented inside the loop.
Next we renormalize the hoppings and the undressed Green's functions, carrying 
the decimation. In the last lines we plot the local density 
of states of site $1$, the local Green's function is given by Eq.~\eqref{eq:G11RGFeff} or
by Eq.~\eqref{3G11} with effective functions. 
The results of few steps are plotted in Fig.~\ref{gr:rgf5917} 
and Fig.~\ref{fig:semiinfingraf}. 

\begin{widetext}
\begin{minipage}{\linewidth}
\begin{algorithm}[H]
\floatname{algorithm}{Source code}
\caption{Semi-infinite chain via surface-bulk recursive Green's function}\label{RGF1D}
\begin{algorithmic}[1]
\begin{Verbatim}[commandchars=\\\{\},frame=none,framesep=1.5ex,framerule=0.8pt,fontsize=\smaller]
\PY{c}{\PYZsh{} Julia programming language: http://julialang.org/}

\PY{k}{using} \PY{n}{PyPlot}                    \PY{c}{\PYZsh{} interface to Matplotlib plotting library}
 
\PY{n}{e0} \PY{o}{=} \PY{l+m+mf}{0.0}                        \PY{c}{\PYZsh{} local site energy}
\PY{n}{eta} \PY{o}{=} \PY{l+m+mf}{1e\PYZhy{}4}                      \PY{c}{\PYZsh{} positive infinitesimal}
\PY{n}{wmin} \PY{o}{=} \PY{o}{\PYZhy{}}\PY{l+m+mf}{2.0}\PY{p}{;} \PY{n}{wmax} \PY{o}{=} \PY{l+m+mf}{2.0}         \PY{c}{\PYZsh{} energy range}
\PY{n}{Nw} \PY{o}{=} \PY{l+m+mi}{1000}                       \PY{c}{\PYZsh{} number of energy points}
\PY{n}{w} \PY{o}{=} \PY{n}{linspace}\PY{p}{(}\PY{n}{wmin}\PY{p}{,}\PY{n}{wmax}\PY{p}{,}\PY{n}{Nw}\PY{p}{)}      \PY{c}{\PYZsh{} linearly spaced vector to store the energies}
\PY{n}{g} \PY{o}{=} \PY{l+m+mf}{1.}\PY{o}{/}\PY{p}{(}\PY{n}{w}\PY{o}{\PYZhy{}}\PY{n}{e0}\PY{o}{+}\PY{n}{eta}\PY{o}{*}\PY{n+nb}{im}\PY{p}{)}            \PY{c}{\PYZsh{} undressed (free site) Green\PYZsq{}s function }
\PY{n}{g10} \PY{o}{=} \PY{n}{g20} \PY{o}{=} \PY{n}{g30} \PY{o}{=} \PY{n}{g}             \PY{c}{\PYZsh{} initialization of undressed GF}
\PY{n}{t} \PY{o}{=} \PY{n}{td} \PY{o}{=} \PY{n}{ones}\PY{p}{(}\PY{n}{Nw}\PY{p}{)}               \PY{c}{\PYZsh{} symmetric real hopping, equal to unity}

\PY{n}{Ndec} \PY{o}{=} \PY{l+m+mi}{16}                       \PY{c}{\PYZsh{} number of decimation iterations}

\PY{k}{for} \PY{n}{i} \PY{k}{in} \PY{l+m+mi}{1}\PY{p}{:}\PY{n}{Ndec}                 \PY{c}{\PYZsh{} Decimation Loop  }

    \PY{n}{g1} \PY{o}{=} \PY{n}{g10}\PY{o}{.}\PY{o}{/}\PY{p}{(}\PY{l+m+mf}{1.0} \PY{o}{\PYZhy{}} \PY{n}{g10}\PY{o}{.}\PY{o}{*}\PY{n}{t}\PY{o}{.}\PY{o}{*}\PY{n}{g20}\PY{o}{.}\PY{o}{*}\PY{n}{td}\PY{p}{)} \PY{c}{\PYZsh{} effective Green\PYZsq{}s function of site 1}
    \PY{n}{g2} \PY{o}{=} \PY{n}{g20}\PY{o}{.}\PY{o}{/}\PY{p}{(}\PY{l+m+mf}{1.0} \PY{o}{\PYZhy{}} \PY{n}{g20}\PY{o}{.}\PY{o}{*}\PY{n}{td}\PY{o}{.}\PY{o}{*}\PY{n}{g20}\PY{o}{.}\PY{o}{*}\PY{n}{t} \PY{o}{\PYZhy{}} \PY{n}{g20}\PY{o}{.}\PY{o}{*}\PY{n}{t}\PY{o}{.}\PY{o}{*}\PY{n}{g20}\PY{o}{.}\PY{o}{*}\PY{n}{td}\PY{p}{)} \PY{c}{\PYZsh{} effective Green\PYZsq{}s function of site 2}
    \PY{n}{g3} \PY{o}{=} \PY{n}{g30}\PY{o}{.}\PY{o}{/}\PY{p}{(}\PY{l+m+mf}{1.0} \PY{o}{\PYZhy{}} \PY{n}{g30}\PY{o}{.}\PY{o}{*}\PY{n}{td}\PY{o}{.}\PY{o}{*}\PY{n}{g20}\PY{o}{.}\PY{o}{*}\PY{n}{t}\PY{p}{)} \PY{c}{\PYZsh{} ./ is an element\PYZhy{}wise division}
			  
    \PY{n}{t} \PY{o}{=} \PY{n}{t}\PY{o}{.}\PY{o}{*}\PY{n}{g20}\PY{o}{.}\PY{o}{*}\PY{n}{t}               \PY{c}{\PYZsh{} Renormalization of the hoppings}
    \PY{n}{td} \PY{o}{=} \PY{n}{td}\PY{o}{.}\PY{o}{*}\PY{n}{g20}\PY{o}{.}\PY{o}{*}\PY{n}{td}            \PY{c}{\PYZsh{} Note that we do not conjugate g20 }

    \PY{n}{g10} \PY{o}{=} \PY{n}{g1}                    \PY{c}{\PYZsh{} Update of the loop variables}
    \PY{n}{g20} \PY{o}{=} \PY{n}{g2}
    \PY{n}{g30} \PY{o}{=} \PY{n}{g3}
\PY{k}{end} 

\PY{n}{G11} \PY{o}{=} \PY{n}{g10}\PY{o}{.}\PY{o}{/}\PY{p}{(}\PY{l+m+mf}{1.0} \PY{o}{\PYZhy{}} \PY{n}{g10}\PY{o}{.}\PY{o}{*}\PY{n}{t}\PY{o}{.}\PY{o}{*}\PY{n}{g20}\PY{o}{.}\PY{o}{*}\PY{n}{td}\PY{o}{.}\PY{o}{/}\PY{p}{(}\PY{l+m+mf}{1.0} \PY{o}{\PYZhy{}} \PY{n}{g20}\PY{o}{.}\PY{o}{*}\PY{n}{t}\PY{o}{.}\PY{o}{*}\PY{n}{g30}\PY{o}{.}\PY{o}{*}\PY{n}{td}\PY{p}{)}\PY{p}{)} \PY{c}{\PYZsh{} final surface Green\PYZsq{}s function of site 1}

\PY{n}{plot}\PY{p}{(}\PY{n}{w}\PY{p}{,} \PY{p}{(}\PY{o}{\PYZhy{}}\PY{l+m+mf}{1.}\PY{o}{/}\PY{n+nb}{pi}\PY{p}{)}\PY{o}{*}\PY{n}{imag}\PY{p}{(}\PY{n}{G11}\PY{p}{)}\PY{p}{)}     \PY{c}{\PYZsh{} Plotting the density of states of the surface site}
\end{Verbatim}

\end{algorithmic}
\end{algorithm}
\end{minipage}
\end{widetext}

In source code \ref{RGF2D}, we have implemented the decimation using the matrix 
forms in \texttt{Julia}. We had to define a vertical and horizontal hopping parameters,
\texttt{tv} and \texttt{tw}, along with hopping matrices \texttt{V} and \texttt{W}.
We now perform an explicit energy and decimation loops, iterating for 1000 energy 
points and 18 decimation steps. Before decimating, we construct a pair of sites, described by the 
dressed function \texttt{gV}, Eq.~\eqref{eq:G1Vmatrix}, coupling two undressed sites.
As shown in Fig.~\ref{fig:laddermap}, we have three effective sites, each one a vertical
pair, and we perform the decimation horizontally, as in the 3-site chain. The decimation 
loop is the same of source code \ref{RGF1D}, except for the fact that we have now a 
hopping matrix \texttt{W}. After the loop, we evaluate the three local functions (as in 
Eq.~\eqref{3G11}, \eqref{3G22} and \eqref{3G33}, but now with effective functions).

\begin{widetext}
\begin{minipage}{\linewidth}
\begin{algorithm}[H]
\floatname{algorithm}{Source code}
\caption{Ladder via surface-bulk recursive Green's function}\label{RGF2D}
\begin{algorithmic}[1]
\begin{Verbatim}[commandchars=\\\{\},frame=none,framesep=1.5ex,framerule=0.8pt,fontsize=\smaller]
\PY{c}{\PYZsh{} Julia programming language version 0.4.2 \PYZhy{} http://julialang.org/}

\PY{k}{using} \PY{n}{PyPlot}                    \PY{c}{\PYZsh{} interface to Matplotlib plotting library}

\PY{n}{e0} \PY{o}{=} \PY{l+m+mf}{0.0}                        \PY{c}{\PYZsh{} local site energy}
\PY{n}{eta} \PY{o}{=} \PY{l+m+mf}{1e\PYZhy{}4}                      \PY{c}{\PYZsh{} positive infinitesimal}
\PY{n}{Ne} \PY{o}{=} \PY{l+m+mi}{1000}                       \PY{c}{\PYZsh{} number of energy points}
\PY{n}{emin} \PY{o}{=} \PY{o}{\PYZhy{}}\PY{l+m+mf}{2.0}\PY{p}{;} \PY{n}{emax} \PY{o}{=} \PY{l+m+mf}{2.0}         \PY{c}{\PYZsh{} energy range}
\PY{n}{envec} \PY{o}{=} \PY{n}{zeros}\PY{p}{(}\PY{n}{Ne}\PY{p}{)}               \PY{c}{\PYZsh{} vector to store the energies}
\PY{n}{tw} \PY{o}{=} \PY{l+m+mf}{1.0}\PY{p}{;}                       \PY{c}{\PYZsh{} hopping between slices}
\PY{n}{tv} \PY{o}{=} \PY{l+m+mf}{1.0}\PY{p}{;} \PY{n}{V} \PY{o}{=} \PY{p}{[}\PY{l+m+mi}{0} \PY{n}{tv}\PY{p}{;} \PY{n}{tv} \PY{l+m+mi}{0}\PY{p}{]}      \PY{c}{\PYZsh{} hopping matrix inside the slice}

\PY{n}{ImG1} \PY{o}{=} \PY{n}{zeros}\PY{p}{(}\PY{n}{Ne}\PY{p}{)}\PY{p}{;} \PY{n}{ImG2} \PY{o}{=} \PY{n}{zeros}\PY{p}{(}\PY{n}{Ne}\PY{p}{)}\PY{p}{;} \PY{n}{ImG3} \PY{o}{=} \PY{n}{zeros}\PY{p}{(}\PY{n}{Ne}\PY{p}{)} \PY{c}{\PYZsh{} global vectors for plotting}
\PY{n}{I} \PY{o}{=} \PY{n}{eye}\PY{p}{(}\PY{l+m+mi}{2}\PY{p}{)}                      \PY{c}{\PYZsh{} eye(n) = nxn identity matrix}

\PY{n}{Ndec} \PY{o}{=} \PY{l+m+mi}{18}                       \PY{c}{\PYZsh{} number of decimation iterations}

\PY{k}{for} \PY{n}{i} \PY{k}{in} \PY{l+m+mi}{1}\PY{p}{:}\PY{n}{Ne}                   \PY{c}{\PYZsh{} Energy loop}

    \PY{n}{en} \PY{o}{=} \PY{n}{emin} \PY{o}{+} \PY{n}{real}\PY{p}{(}\PY{n}{i}\PY{o}{\PYZhy{}}\PY{l+m+mi}{1}\PY{p}{)}\PY{o}{*}\PY{p}{(}\PY{n}{emax}\PY{o}{\PYZhy{}}\PY{n}{emin}\PY{p}{)}\PY{o}{/}\PY{p}{(}\PY{n}{Ne}\PY{o}{\PYZhy{}}\PY{l+m+mi}{1}\PY{p}{)} \PY{c}{\PYZsh{} energy \PYZhy{} real(n) is a conversion to float}
    \PY{n}{W} \PY{o}{=} \PY{p}{[}\PY{n}{tw} \PY{l+m+mi}{0}\PY{p}{;} \PY{l+m+mi}{0} \PY{n}{tw}\PY{p}{]}            \PY{c}{\PYZsh{} hopping matrix \PYZhy{} between slices}

    \PY{n}{g} \PY{o}{=} \PY{p}{(}\PY{l+m+mf}{1.}\PY{o}{/}\PY{p}{(}\PY{n}{en}\PY{o}{\PYZhy{}}\PY{n}{e0}\PY{o}{+}\PY{n}{eta}\PY{o}{*}\PY{n+nb}{im}\PY{p}{)}\PY{p}{)}\PY{o}{*}\PY{n}{I}   \PY{c}{\PYZsh{} undressed Green\PYZsq{}s function of a site}
    \PY{n}{gV} \PY{o}{=} \PY{n}{inv}\PY{p}{(}\PY{n}{I} \PY{o}{\PYZhy{}} \PY{n}{g}\PY{o}{*}\PY{n}{V}\PY{p}{)}\PY{o}{*}\PY{n}{g}         \PY{c}{\PYZsh{} Green\PYZsq{}s function of a vertically coupled pair of sites}
    \PY{n}{g1} \PY{o}{=} \PY{n}{gV}\PY{p}{;} \PY{n}{g2} \PY{o}{=} \PY{n}{gV}\PY{p}{;} \PY{n}{g3} \PY{o}{=} \PY{n}{gV}   \PY{c}{\PYZsh{} initialization of three isolated slices}

    \PY{k}{for} \PY{n}{j} \PY{k}{in} \PY{l+m+mi}{1}\PY{p}{:}\PY{n}{Ndec}             \PY{c}{\PYZsh{} Decimation Loop in the horizontal direction }

        \PY{n}{g1n} \PY{o}{=} \PY{n}{inv}\PY{p}{(}\PY{n}{I} \PY{o}{\PYZhy{}} \PY{n}{g1}\PY{o}{*}\PY{n}{W}\PY{o}{*}\PY{n}{g2}\PY{o}{*}\PY{n}{W}\PY{p}{)}\PY{o}{*}\PY{n}{g1}  \PY{c}{\PYZsh{} effective auxiliary Green\PYZsq{}s functions}
        \PY{n}{g2n} \PY{o}{=} \PY{n}{inv}\PY{p}{(}\PY{n}{I} \PY{o}{\PYZhy{}} \PY{p}{(}\PY{n}{g2}\PY{o}{*}\PY{n}{W}\PY{o}{.}\PY{o}{\PYZsq{}}\PY{o}{*}\PY{n}{g2}\PY{o}{*}\PY{n}{W}\PY{p}{)} \PY{o}{\PYZhy{}} \PY{p}{(}\PY{n}{g2}\PY{o}{*}\PY{n}{W}\PY{o}{*}\PY{n}{g2}\PY{o}{*}\PY{n}{W}\PY{o}{.}\PY{o}{\PYZsq{}}\PY{p}{)}\PY{p}{)}\PY{o}{*}\PY{n}{g2} 
        \PY{n}{g3n} \PY{o}{=} \PY{n}{inv}\PY{p}{(}\PY{n}{I} \PY{o}{\PYZhy{}} \PY{n}{g3}\PY{o}{*}\PY{n}{W}\PY{o}{.}\PY{o}{\PYZsq{}}\PY{o}{*}\PY{n}{g2}\PY{o}{*}\PY{n}{W}\PY{p}{)}\PY{o}{*}\PY{n}{g3} 

        \PY{n}{W} \PY{o}{=} \PY{n}{W}\PY{o}{*}\PY{n}{g2}\PY{o}{*}\PY{n}{W}              \PY{c}{\PYZsh{} effective hopping}
        \PY{n}{g1} \PY{o}{=} \PY{n}{g1n}                \PY{c}{\PYZsh{} update of the variables }
        \PY{n}{g2} \PY{o}{=} \PY{n}{g2n} 
        \PY{n}{g3} \PY{o}{=} \PY{n}{g3n} 
    \PY{k}{end} 

    \PY{c}{\PYZsh{} local Green\PYZsq{}s functions}
    \PY{n}{G1} \PY{o}{=} \PY{n}{inv}\PY{p}{(}\PY{n}{I} \PY{o}{\PYZhy{}} \PY{p}{(}\PY{n}{g1}\PY{o}{*}\PY{n}{W}\PY{o}{*}\PY{n}{g2}\PY{o}{*}\PY{n}{W}\PY{o}{.}\PY{o}{\PYZsq{}}\PY{p}{)}\PY{o}{*}\PY{n}{inv}\PY{p}{(}\PY{n}{I} \PY{o}{\PYZhy{}} \PY{n}{g2}\PY{o}{*}\PY{n}{W}\PY{o}{*}\PY{n}{g3}\PY{o}{*}\PY{n}{W}\PY{o}{.}\PY{o}{\PYZsq{}}\PY{p}{)}\PY{p}{)}\PY{o}{*}\PY{n}{g1}     
    \PY{n}{G2} \PY{o}{=} \PY{n}{g2}
    \PY{n}{G3} \PY{o}{=} \PY{n}{inv}\PY{p}{(}\PY{n}{I} \PY{o}{\PYZhy{}} \PY{p}{(}\PY{n}{g3}\PY{o}{*}\PY{n}{W}\PY{o}{.}\PY{o}{\PYZsq{}}\PY{o}{*}\PY{n}{g2}\PY{o}{*}\PY{n}{W}\PY{p}{)}\PY{o}{*}\PY{n}{inv}\PY{p}{(}\PY{n}{I} \PY{o}{\PYZhy{}} \PY{n}{g2}\PY{o}{*}\PY{n}{W}\PY{o}{.}\PY{o}{\PYZsq{}}\PY{o}{*}\PY{n}{g1}\PY{o}{*}\PY{n}{W}\PY{p}{)}\PY{p}{)}\PY{o}{*}\PY{n}{g3} 

    \PY{n}{envec}\PY{p}{[}\PY{n}{i}\PY{p}{]} \PY{o}{=} \PY{n}{en}               \PY{c}{\PYZsh{} storing the energy mesh}

    \PY{n}{ImG1}\PY{p}{[}\PY{n}{i}\PY{p}{]} \PY{o}{=} \PY{n}{imag}\PY{p}{(}\PY{n}{G1}\PY{p}{[}\PY{l+m+mi}{1}\PY{p}{,}\PY{l+m+mi}{1}\PY{p}{]}\PY{p}{)}     \PY{c}{\PYZsh{} storing the imaginary part }
    \PY{n}{ImG2}\PY{p}{[}\PY{n}{i}\PY{p}{]} \PY{o}{=} \PY{n}{imag}\PY{p}{(}\PY{n}{G2}\PY{p}{[}\PY{l+m+mi}{1}\PY{p}{,}\PY{l+m+mi}{1}\PY{p}{]}\PY{p}{)} 
    \PY{n}{ImG3}\PY{p}{[}\PY{n}{i}\PY{p}{]} \PY{o}{=} \PY{n}{imag}\PY{p}{(}\PY{n}{G3}\PY{p}{[}\PY{l+m+mi}{1}\PY{p}{,}\PY{l+m+mi}{1}\PY{p}{]}\PY{p}{)} 
\PY{k}{end}

\PY{c}{\PYZsh{} Plotting}

\PY{n}{subplot}\PY{p}{(}\PY{l+m+mi}{211}\PY{p}{)}                    \PY{c}{\PYZsh{} Create the first plot of a 2x1 group of subplots}
\PY{n}{plot}\PY{p}{(}\PY{n}{envec}\PY{p}{,} \PY{p}{(}\PY{o}{\PYZhy{}}\PY{l+m+mf}{1.}\PY{o}{/}\PY{n+nb}{pi}\PY{p}{)}\PY{o}{.}\PY{o}{*}\PY{n}{ImG1}\PY{p}{,} \PY{n}{linewidth}\PY{o}{=}\PY{l+m+mf}{3.0}\PY{p}{,} \PY{n}{label}\PY{o}{=}\PY{l+s}{L\PYZdq{}}\PY{l+s}{G\PYZus{}1(1,1)}\PY{l+s}{\PYZdq{}}\PY{p}{,} \PY{n}{color}\PY{o}{=}\PY{l+s}{\PYZdq{}}\PY{l+s}{\PYZsh{}85beff}\PY{l+s}{\PYZdq{}}\PY{p}{)}
\PY{n}{plot}\PY{p}{(}\PY{n}{envec}\PY{p}{,} \PY{p}{(}\PY{o}{\PYZhy{}}\PY{l+m+mf}{1.}\PY{o}{/}\PY{n+nb}{pi}\PY{p}{)}\PY{o}{.}\PY{o}{*}\PY{n}{ImG3}\PY{p}{,} \PY{n}{linestyle}\PY{o}{=}\PY{l+s}{\PYZdq{}}\PY{l+s}{\PYZhy{}\PYZhy{}}\PY{l+s}{\PYZdq{}}\PY{p}{,} \PY{n}{linewidth}\PY{o}{=}\PY{l+m+mf}{2.0}\PY{p}{,} \PY{n}{label}\PY{o}{=}\PY{l+s}{L\PYZdq{}}\PY{l+s}{G\PYZus{}3(1,1)}\PY{l+s}{\PYZdq{}}\PY{p}{)} 
\PY{n}{legend}\PY{p}{(}\PY{n}{loc}\PY{o}{=}\PY{l+s}{\PYZdq{}}\PY{l+s}{upper right}\PY{l+s}{\PYZdq{}}\PY{p}{,}\PY{n}{fancybox}\PY{o}{=}\PY{l+s}{\PYZdq{}}\PY{l+s}{true}\PY{l+s}{\PYZdq{}}\PY{p}{)}
\PY{n}{ylabel}\PY{p}{(}\PY{l+s}{\PYZdq{}}\PY{l+s}{Density of states}\PY{l+s}{\PYZdq{}}\PY{p}{,} \PY{n}{fontsize}\PY{o}{=}\PY{l+m+mi}{16}\PY{p}{)}

\PY{n}{subplot}\PY{p}{(}\PY{l+m+mi}{212}\PY{p}{)}                    \PY{c}{\PYZsh{} Create the 2nd plot of a 2x1 group of subplots}
\PY{n}{plot}\PY{p}{(}\PY{n}{envec}\PY{p}{,} \PY{p}{(}\PY{o}{\PYZhy{}}\PY{l+m+mf}{1.}\PY{o}{/}\PY{n+nb}{pi}\PY{p}{)}\PY{o}{.}\PY{o}{*}\PY{n}{ImG2}\PY{p}{,} \PY{n}{label}\PY{o}{=}\PY{l+s}{L\PYZdq{}}\PY{l+s}{G\PYZus{}2(1,1)}\PY{l+s}{\PYZdq{}}\PY{p}{,} \PY{n}{linewidth}\PY{o}{=}\PY{l+m+mf}{3.0}\PY{p}{,} \PY{n}{color}\PY{o}{=}\PY{l+s}{\PYZdq{}}\PY{l+s}{g}\PY{l+s}{\PYZdq{}}\PY{p}{)} 
\PY{n}{legend}\PY{p}{(}\PY{n}{loc}\PY{o}{=}\PY{l+s}{\PYZdq{}}\PY{l+s}{upper right}\PY{l+s}{\PYZdq{}}\PY{p}{,}\PY{n}{fancybox}\PY{o}{=}\PY{l+s}{\PYZdq{}}\PY{l+s}{true}\PY{l+s}{\PYZdq{}}\PY{p}{)}
\PY{n}{xlabel}\PY{p}{(}\PY{l+s}{L\PYZdq{}}\PY{l+s+si}{\PYZdl{}}\PY{l+s}{(}\PY{l+s}{\PYZbs{}}\PY{l+s}{omega\PYZhy{}}\PY{l+s}{\PYZbs{}}\PY{l+s}{epsilon\PYZus{}0)/t }\PY{l+s+si}{\PYZdl{}}\PY{l+s}{\PYZdq{}}\PY{p}{,} \PY{n}{fontsize}\PY{o}{=}\PY{l+m+mi}{20}\PY{p}{)}
\PY{n}{ylabel}\PY{p}{(}\PY{l+s}{\PYZdq{}}\PY{l+s}{Density of states}\PY{l+s}{\PYZdq{}}\PY{p}{,} \PY{n}{fontsize}\PY{o}{=}\PY{l+m+mi}{16}\PY{p}{)}
\end{Verbatim}

\end{algorithmic}
\end{algorithm}
\end{minipage}
\end{widetext}

\end{document}